\documentclass[prd,aps,a4paper,floatfix,amsmath,amssymb,twocolumn,nofootinbib]{revtex4-1}
\usepackage{graphicx,xcolor,array,float,fullpage}
\usepackage[pdftex,breaklinks,colorlinks,linkcolor=blue,citecolor=teal,anchorcolor=red,urlcolor=cyan]{hyperref}

\usepackage[caption=false]{subfig}

\begin{document}


\title{Critical phenomena in gravitational collapse with competing
  scalar field and gravitational waves, in 4+1 dimensions}
\author{Bernardo Porto Veronese} \affiliation{\'Ecole
  Polytechnique, Route de Saclay, 91128 Palaiseau Cedex, France}
\author{Carsten Gundlach} \affiliation{Mathematical Sciences,
  University of Southampton, Southampton SO17 1BJ, United Kingdom}
\date{17 September 2022}


\begin{abstract}

In the gravitational collapse of matter beyond spherical symmetry,
gravitational waves are necessarily present. On the other hand,
gravitational waves can collapse to a black hole even without
matter. One might therefore wonder how the interaction and competition
between the matter fields and gravitational waves affects critical
phenomena at the threshold of black hole formation. As a toy model for
this, we study the threshold of black-hole formation in 4+1
dimensions, where we add a massless minimally coupled scalar matter
field to the gravitational wave ansatz of Biz\'on, Chmaj and Schmidt
(in a nutshell, Bianchi~IX on
$S^3\times\text{radius}\times\text{time}$). In order to find a stable
discretisation of the equation governing the gravitational waves in
4+1 physical dimensions, which has the same principal part as the
spherical wave equation in 9+1 dimensions, we first revisit the
problem of critical spherical scalar field collapse in $n+2$
dimensions with large $n$. Returning to the main problem, we find
numerically that weak gravitational wave perturbations of the scalar
field critical solution decay, while weak scalar perturbations of the
gravitational wave critical solution also decay. A dynamical systems
picture then suggests the existence of a codimension-two attractor. We
find numerical evidence for this attractor by evolving mixed initial
data and fine-tuning both an overall amplitude and the relative
strength of the two fields.

\end{abstract}

\maketitle

\tableofcontents


\section{Introduction}


In many self-gravitating systems that are exactly scale-invariant, or
asymptotically scale-invariant on small scales, numerical time
evolutions of regular, finite mass initial data show that data which
are fine-tuned more and more closely to the threshold of collapse, but
otherwise generic, evolve into arbitrarily small black holes  on the
supercritical side of the threshold, arbitrarily large curvature
before dispersion on the subcritical side. This is known as ``type II
critical phenomena in gravitational collapse'', see \cite{critreview}
for a review.

The near-critical time evolutions go through a universal 
  codimension-one attractor that is self-similar (or asymptotically
self-similar on small scales), and which itself has a naked
singularity,  called the ``critical solution''. In the limit of
perfect fine-tuning of any one parameter of the initial data to the
collapse threshold, the time evolution  approaches but never
leaves the critical solution, and so a naked singularity is generated
in the time evolution of a codimension-one set of otherwise generic
initial data.

This is well established numerically, and well understood
mathematically, for a number of Einstein-matter systems in spherical
symmetry, see \cite{critreview}. Moreover, for at least some of these
systems, type~II critical collapse is stable under small but finite
non-spherical perturbations
\cite{Baumgartescalar,BaumgarteGundlach}. Going beyond spherical
symmetry is interesting for at least two reasons: it allows for
angular momentum, and for gravitational collapse in vacuum.

Vacuum critical collapse is of interest as it is not tied to a
particular choice of matter. However, fine-tuning to the threshold of
collapse in vacuum gravity has proved numerically very difficult even
in twist-free axisymmetry, see
\cite{Suarez_etal_2022,Ledvinka_Khirnov_2021} for the current state of
the art. As a stepping stone from vacuum, critical collapse has been
investigated in twist-free axisymmetry with matter, in particular a
perfect fluid \cite{BaumgarteGundlach} and electromagnetic radiation
\cite{Baumgarte_Gundlach_Hilditch_2019}. However, in going beyond
spherical symmetry, the moving matter necessarily also creates
gravitational waves. In the critical collapse of axisymmetric
electromagnetic waves, an approximately discretely self-similar (from
now on, DSS) critical solution was observed, but with
scale-periodicity less regular than that observed in spherical scalar
field collapse \cite{Baumgarte_Gundlach_Hilditch_2019}. It was
conjectured that this is due to the effect of strong gravitational
waves.

As a spherically symmetric toy model for this interaction of matter
and gravitational waves,  one of us with collaborators
\cite{Gundlach_Baumgarte_Hilditch_2019} investigated critical collapse
with two massless matter fields, a Yang-Mills (from now on, YM) and a
scalar field. They found the well-known critical solutions for pure YM
and pure scalar field matter. Perturbing pure initial data with an
infinitesimal amount of the other type of matter, they established
that weak YM perturbations of the scalar field critical solution
decay, but that weak scalar perturbations of the YM critical solution
grow.

Setting up mixed initial data with different ratios, and fine-tuning
again to the black-hole threshold, they found a mixed-field critical
solution that starts as a growing perturbation of the pure YM critical
solution (at large scales) and ends as a decaying perturbation of the
pure scalar critical solution (at small scales). This solution changes
its matter content from pure YM to pure scalar field on the fly, while
remaining very compact (with $2M/R\sim 0.5$) and approximately
DSS, with the approximate log-scale period $\Delta$ changing from the
YM to the scalar field value.

Here we investigate another toy model, where the two interacting
fields truly are gravitational waves and a massless matter
field. Biz\'on, Chmaj and Schmidt \cite{Bizon_Chmaj_Schmidt_2005}
proposed an ansatz in 4+1 spacetime dimensions on
the manifold $S^3\times({\Bbb R}\times{\Bbb R}^+)$, where the metric
on the factor $S^3$ is homogeneous but anisotropic, namely, it is of
Bianchi type~IX. Here all metric variables depend only on time and
radius, even though the spacetime is vacuum. (This can be generalised
to higher odd-dimensional spheres). To this system we simply add a
 homogeneous massless minimally coupled scalar field $\Psi$.

We thus have a toy model for matter coupled to gravitational waves,
but where all fields depend only on radius and time, so that numerical
time evolutions are cheap. Besides the unphysical dimensions, the
major shortcoming of this model is that the scalar field cannot create
gravitational waves if they are absent initially -- we shall discuss
this in more detail below.

The field equations for the scalar field and the gravitational waves
are essentially spherical wave equations, in the physical 4+1
dimensions for the scalar field $\Psi$, but effectively in 9+1
dimensions for the gravitational wave variable $b$. As is well-known,
such spherical wave equations are numerically difficult in high
dimensions. It turns out the methods that work well in 3+1 dimensions
stretch to 4+1 but not to 9+1 dimensions. As a stepping stone, we were
therefore forced to revisit the problem of critical collapse of a
spherically symmetric scalar field in high dimensions. In
Appendix~\ref{appendix:scalarfield} we re-derive and modify the method
of \cite{Bland_Preston_Becker_Kunstatter_Husain_2005} and present
successful tests in critical scalar field collapse in 9+1 (physical)
dimensions.

In Sec.~\ref{section:numericalmethod} we present our discretisation of
the field equations, using the methods of
Appendix~\ref{appendix:scalarfield} for the field $b$, and in
Sec.~\ref{section:similaritysolutions} the similarity coordinates that
we use to display the approximate self-similarity of near-critical
time evolutions. Sec.~\ref{section:results} contains our numerical
results, and Sec.~\ref{section:conclusions} our conclusions.


\section{Metric ansatz and field equations}
\label{sec:metric-and-equations}


We make the Bianchi~IX ansatz of \cite{Bizon_Chmaj_Schmidt_2005},
restricting to the biaxial case. We introduce null coordinates adapted
to the Bianchi~IX symmetry $(u,x, \theta, \varphi, \psi)$, in terms of
which the line element becomes
\begin{eqnarray}
\label{eq:uxmetric}
ds^2&=&- 2 G\, du\, dx - H du^2 + \frac{1}{4} R^2 \Big( e^{2 B}\,  d\theta^2
\nonumber \\
&& +  
 (e^{2B} \cos^2 \theta + e^{-4 B} \sin^2 \theta)\, d\varphi^2
\nonumber \\
&& - 2 e^{-4B} \sin \theta\, d\varphi\, d\theta +  e^{-4 B}\, d\psi^2 \Big).    
\end{eqnarray}
The coordinate $u$ is null, and the tangent vector to  the
  affinely parameterised outgoing null geodesics ruling the
surfaces of constant $u$ is $U^a:=-\nabla^a
u=G^{-1}(\partial_x)^a$. Here $G$, $H$, $R$ and $B$ are functions of
$u$ and $x$ only. We also introduce the derivative operator
\begin{equation}
    \Xi := \partial_u - \frac{H}{2G} \partial_x,
\end{equation}
which is tangential to the ingoing null rays emanating from the
3-surfaces of constant $u$ and $x$. In the special case $H=0$, $x$ is
also a null coordinate and $\Xi=\partial_u$.

We fix the remaining coordinate freedom in the ansatz
(\ref{eq:uxmetric}) by imposing
\begin{eqnarray}
\label{myH}
\frac{H}{2G} &=& \left(1-\frac{x}{x_0} \right)\frac{1}{2R_{,x}(u, 0)},
\\
\label{gatcentre}
G(u,0)&=&R_{,x}(u, 0), \\
\label{Rinitial}
R(0,x)&=&\frac{x}{2}.
\end{eqnarray}
This puts the centre $R=0$ at $x=0$, makes $u$ the proper time there,
and makes $x=x_0$ an ingoing null surface. More generally, surfaces of
constant $x$ are timelike for $0\le x<x_0$ and spacelike for
$x>x_0$. In particular, choosing the outer boundary of our numerical
domain at $x=x_\text{max}>x_0$ means that this boundary is future
spacelike and no boundary condition is required.

Moreover, if $x_0$ is chosen so that the ingoing lightcone $x=x_0$ is
approximately the past lightcone of the accumulation point $(u_*,0)$
of scale echoes of an (approximately) self-similar spacetime, our
coordinate system automatically zooms in on this point, giving us good
resolution in critical collapse without the need for explicit mesh
refinement. 

Our coordinate $x$ can be related to an ingoing
null coordinate $v$ by
\begin{equation}
v(u,x)=-f(u)\left(1-\frac{x}{x_0}\right),
\end{equation}
where 
\begin{equation}
f(u)=\exp\left[-x_0\int_0^u \frac{du'}{2R_{,x}(u',0)}\right].
\end{equation}
$v$ is an increasing linear function of $x$, such that $v=0$ is mapped
to $x=x_0$. Our coordinate system can therefore be thought of as a
continuous version of Garfinkle's algorithm \cite{Garfinkle_1995},
which rescales $v$ linearly in what in our notation is called $x$, but
by interpolation at discrete moments of time $u$, rather than the
continuous use of a radial shift vector. We had previously used
Garfinkle's method in \cite{Baumgarte_Gundlach_Hilditch_2019}, and for
that problem our new algorithm gives the same accuracy and run
times. We have made the change here as it simplifies convergence
testing. Both algorithms require a good choice of, in our notation,
$x_0$ in order to make the coordinate system zoom in on the
accumulation point of critical collapse.

To regularise the field equations, we redefine two of the metric
coefficients as
\begin{equation}
\label{eq:small-b}
B=:R^2b
\end{equation}
and
\begin{equation}
G=:R_{,x}g.
\end{equation}

There are four algebraically independent components of the Einstein
equations
\begin{equation}
R_{ab}=8\pi \nabla_a \Psi\nabla_b \Psi. 
\end{equation}
(We work in units where $G=c=1$.) From these, we select one
which is an ordinary differential equation for $g$
on the slices of constant $u$, and two which are wave quations for
$R$ and $b$. The remaining Einstein equation is then redundant. We
also have a wave equation for the matter field $\Psi$.

The four field equations thus obtained can be arranged in the
following hierarchy:
\begin{eqnarray}
\label{Dlng}
  \mathcal{D}(\ln g)&=& \frac{8\pi R}{3} (\mathcal{D} \Psi )^2
  \nonumber \\ 
&& + 2 R^3  (\mathcal{D}b + 2 R b)^2, \\
\label{DR2Ru}
  \mathcal{D}(R^2 \Xi R) &=& \frac{g R}{3} (1-4 e^{6bR^2}) e^{-8bR^2},
  \\
\label{DR32Psiu}
  \mathcal{D}(R^{3/2} \Xi \Psi)&=& -\frac{3}{2} \Xi R R^{1/2}
  (\mathcal{D} \Psi),
\\
\label{DR72bu}
 \mathcal{D}(R^{7/2}\Xi b) &=& \frac{2}{3} g R^{-1/2} e^{-8bR^2}
 \Bigl(1-e^{6bR^2} \nonumber\\ &&
+ bR^2 (4 e^{6bR^2}-1)\Bigr) \nonumber\\ && -4b R^{3/2} \Xi R
 - \frac{7}{2} R^{5/2} \Xi R \mathcal{D}b.
\end{eqnarray}
Here
\begin{equation}
\mathcal{D}f:={f_{,x}\over R_{,x}},
\end{equation}
so that ${\cal D}$ is $d/dR$ along the null geodesics ruling the
slices of constant $u$. Note that these equations do not explicitly
contain $H$. Rather, $H$ can be chosen freely [we choose (\ref{myH})],
and appears only when we use 
\begin{equation}
\Psi_{,u}=\Xi\Psi+{H\over 2G}\Psi_{,x}
\end{equation}
in order to advance $\Psi$ in $u$, and similarly for $b$ and $R$.

Eqs.~(\ref{Dlng}-\ref{DR72bu}) can be solved for $g$, $\Xi R$,
$\Xi \Psi$ and $\Xi b$ in the above order by the integration
\begin{equation}
\mathcal{I}f:=\int f R_{,x}\,dx=\int f\,dR
\end{equation}
along the outgoing null geodesics, labelled by constant
$(u,\theta,\varphi, \psi)$, starting the integration from the centre
$R=0$. Because of factors of $R$, three of the startup conditions are
selected by regularity at $R=0$. The fourth startup condition at $R=0$
is the gauge choice $g=1$, equivalent to (\ref{gatcentre}) above.

This selection and hierarchical arrangement of the field equations
closely resembles the form of the field equations for the spherical
scalar field and YM field of \cite{Gundlach_Baumgarte_Hilditch_2019},
with $\partial_u$ replaced by its generalisation $\Xi$. Somewhat less
closely, it also resembles the formulation for the spherical scalar
field of
\cite{Goldwirth_Piran_1987,Gundlach_Price_Pullin_1994,Garfinkle_1995}
(but with ${\cal D}$ and $\Xi$ applied to $\Psi$ in the opposite
order), and the scheme of \cite{Gomez_Papadopoulos_Winicour_1994} for
the vacuum Einstein equations on null cones with a regular vertex (but
in terms of null coordinates $u$ and $x$, rather than Bondi
coordinates $u$ and $R$).

In analogy with the field redefinitions made in
\cite{Bland_Preston_Becker_Kunstatter_Husain_2005} (see also
  Appendix~\ref{appendix:scalarfield}) we replace $b$ as an evolved
  variable by
\begin{equation}
\label{eq:chi-def}
    \chi: = b + \frac{2}{7}R\mathcal{D}b,
\end{equation}
from which we can reconstruct $b$ as 
\begin{equation}
\label{eq:b-from-chi}
b = \frac{1}{R^{7/2}}\int_0^{R} \chi\, d(\tilde{R}^{7/2}).
\end{equation}
The computation of $b$ from $\chi$ is more stable numerically if
we integrate \eqref{eq:b-from-chi} by parts, giving us
\begin{equation}
\label{eq:b-from-chi-integated-by-parts}
    b = \chi - \frac{2}{9}\frac{1}{R^{7/2}}\int_0^R \mathcal{D}\chi
    \, d(\tilde{R}^{9/2}).
\end{equation}
The second term on the right-hand side of
Eq.~\eqref{eq:b-from-chi-integated-by-parts} is $O(R)$ near the
origin, and thus generates less error from finite differencing than
the original integral in Eq.~\eqref{eq:b-from-chi}, which is $O(1)$
there. 

The evolution equation for $\chi$ is
\begin{eqnarray}
\label{eq:chi-evol-three-terms}
   \Xi \chi = &&\frac{4}{21}\frac{g}{R^{3}}\Gamma
    (bR^2)-\frac{8}{7}\frac{b}{R}\left(\Xi
    R+\frac{g}{2}\right)\nonumber\\ &&- \frac{(\chi - b)}{2R}
    \left[\Xi R + \frac{2g}{3}(1-4e^{6bR^2})e^{-8bR^2}\right], \nonumber\\
\end{eqnarray}
where $\Gamma(x):=3x+e^{-8x}(1-e^{6x}+x(4e^{6x}-1))$. Its series
expansion is $\Gamma(x)=30x^2+O(x^3)$, and so the leading $b^2R^4$
term near the origin cancels the first denominator of
Eq.~\eqref{eq:chi-evol-three-terms}. 

Furthermore, the expression $\Xi R+g/2$, which appears in the second
term of Eq.~\eqref{eq:chi-evol-three-terms}, is $O(R^2)$ near the
origin.  This cancels the denominator of the second term.  To see this
explicitly, manifestly cancel the $O(1)$ and $O(R)$ contributions in
$\Xi R$ and $g/2$ by integrating Eq.~\eqref{DR2Ru} by parts, giving us
\begin{eqnarray}
    && \Xi R-\frac{g}{6}(1-4e^{6bR^2})e^{-8bR^2} \nonumber \\
&=&-\frac{1}{R^2}\int_0^R
    \frac{g\tilde{R}^3}{6}\bigg[ \frac{8 \pi}{3} (\mathcal{D}\Psi)^2
      \nonumber \\
&& +2\tilde{R}^2(\tilde{R}\mathcal{D}b+2b)^2 \nonumber \\
&& +
      8(\tilde{R}\mathcal{D}b+2b)(1-e^{-6b\tilde{R}^2})e^{-2b\tilde{R}^2}\bigg]
    d\tilde{R}.
\label{eqnnotused}
\end{eqnarray}
The left hand side equals $\Xi R + g/2 + O(R^4)$, and from the
regularity of $\Psi$ and $b$ the integral on the right hand side is
$O(R^4)$. [We do not use Eq.~(\ref{eqnnotused}) in our code. It
    is given here just to show that
    Eq.~\eqref{eq:chi-evol-three-terms} is explicitly regular.]

Finally, the regularity of the last term on the right hand side of
Eq.~\eqref{eq:chi-evol-three-terms} follows from the definition of
$\chi$, Eq.~\eqref{eq:chi-def}.

We now introduce some diagnostics. We define the Misner-Sharp-like
quasilocal mass function $M(u,x)$, and the related compactness ${\cal
  C}$, by
\begin{equation} \label{compactness}
\mathcal{C}:={M\over R^2}:=1-\nabla_aR\nabla^aR=1+2{\Xi R \over g}.
\end{equation}
In spherical symmetry, a marginally outer-trapped surface (from now on
also referred to as an apparent horizon), occurs where
$\mathcal{C}=1$, but our formulation of the Einstein equations does
not allow us to reach this. Rather, we take $\mathcal{C}\to 1$ as an
approximate criterion for apparent horizon formation.

For the diagnosis of subcritical scaling we introduce the
curvature-like quantities
\begin{eqnarray}
R_\Psi&:=&{R^a}_a= 8\pi \nabla^a \Psi \nabla_a \Psi= -{16\pi \over
  g}\Xi \Psi {\cal D}\Psi, \nonumber \\ \\
R_B&:=& 6\nabla^a B \nabla_a B=  -{12\over g}\Xi B {\cal D}B \nonumber
\\
&=&-{12\over g}(R^2 \Xi b +2R\Xi R b)(R^2{\cal D}b+2Rb). \nonumber \\
\end{eqnarray}
$R_\Psi$ is actually the Ricci scalar, which is determined by $\Psi$
alone, while $B$ does not contribute to the Ricci tensor at all.
However, $\Psi$ and $B$ appear in a similar manner both in the
Einstein equation for $g$, namely
\begin{equation}
 \mathcal{D}(\ln g)= {2R\over3}\left(4\pi(\mathcal{D}\Psi)^2 +
 3(\mathcal{D}B)^2\right)
\end{equation}
[compare Eq.~(\ref{Dlng})],
and in the mass aspect, namely
\begin{eqnarray}
{\cal D}M&=&{2R\over3}\Bigl[
3+e^{-8B}-4e^{-2B}\nonumber \\
&&+(R^2-M)\left(4\pi(\mathcal{D}\Psi)^2+3(\mathcal{D}B)^2
 \right)\Bigr].
\end{eqnarray}
We have adjusted the overall constant factor in the definition of
$R_B$ to reflect this. Note that $R_\Psi$ is nonzero at the
  centre, whereas $R_B\sim R^2$ vanishes there.

Even though B represents genuine gravitational waves, their
polarisation is in the angular, homogeneous, directions
$(\theta,\varphi,\psi)$, while the scalar field depends only on the
orthogonal directions $(u,x)$. Therefore the scalar matter field
cannot create gravitational waves if they are absent initially, in
contrast to the case of electromagnetic waves, or a non-spherical
scalar field or fluid, in 3+1 dimensions. In this respect, the system
looks mathematically more similar to that of
\cite{Gundlach_Baumgarte_Hilditch_2019} (two matter fields coupled to
each other only through the metric) than to, say, a massless scalar
field minimally coupled to gravity in axisymmetry.


\section{Numerical method}
\label{section:numericalmethod}


Our numerical implementation is an adaptation of that of
\cite{Gundlach_Baumgarte_Hilditch_2019}. We represent our fields on a
grid with $N_x = 600$ equally spaced points in $x$, and numerically
advance in the retarded time $u$. We set $x_i=i\Delta x$ for $1\le
1\le N_x$, with $x=0$ not on the grid. We extrapolate to $x=0$ where
needed, and for output only, but we use the assumption that
  $R=0$ there in our boundary conditions.

At every time step, we solve for $b$, $g$ and the ingoing null
derivatives $\Xi R$, $\Xi \Psi$ and $\Xi \chi$ from
(\ref{eq:b-from-chi-integated-by-parts}), the integrated versions of
Eqs.~(\ref{Dlng}-\ref{DR32Psiu}), and (\ref{eq:chi-evol-three-terms}),
in this order. We then evolve $R$, $\Psi$ and $\chi$ from $u$ to
$u+\Delta u$ using a second-order Runge-Kutta method.  We use the
heuristic timestep criterion
\begin{equation}
\label{Cdef}
|\Xi R|\Delta u\le C R_{,x}\Delta x,
\end{equation}
implemented as
\begin{equation}
\Delta u =C \min_i {2(R_{i}-R_{i-1})\over \max(\Xi R_i,\Xi R_{i-1})}.
\end{equation}
$C$ is a dimensionless factor of order unity, as in
\cite{Gundlach_Baumgarte_Hilditch_2019}. We use $C=0.1$ throughout.

To start up the integration of the Einstein equations, we make the
least-squares fit $\Psi\simeq \Psi_0+\Psi_1R+\mathcal{O}(R^2)$ to the
two innermost grid points. We then substitute these expansions into
the integral expressions for $g$, $\Xi R$, $\Xi \Psi$ and $\Xi \chi$,
obtaining
\begin{eqnarray}
g &=&1+\frac{4\pi \Psi_1^2}{3} R^2  +\mathcal{O}(R^3), \\
\Xi R &=& - \frac{1}{2} - \frac{\pi \Psi_1^2}{3} R^2 + \mathcal{O}(R^3), \\
\label{eq:xi-psi-expansion}
\Xi \Psi &=&{\Psi_1\over 2}+\mathcal{O}(R).
\end{eqnarray}
These expansions are used at the first grid point to start up the
integrations for $g, \Xi R$ and $\Xi \Psi$. No linear expansion is
required for $\Xi \chi$ as there is no integral. 

The derivative $\mathcal{D}$ is discretised by symmetric finite
differencing with respect to $R$:
\begin{equation}
    (\mathcal{D}\Psi)_i = \frac{\Psi_i - \Psi_{i-1}}{R_i - R_{i-1}},
\end{equation} 
and likewise for $\chi$ and $b$. Indicating by 
\begin{equation}
    \bar{\Psi}_i = \frac{\Psi_i+\Psi_{i-1}}{2}
\end{equation} 
the numerical approximation of $\Psi$ in the midpoint of the $i$-th
grid cell (and likewise for other quantities), the integrals over the
grid points $1, \ldots, j$ are then discretised using the midpoint
rule:
\begin{eqnarray}
    &&	\int_{R_1}^{R_j} f(\Psi,\mathcal{D}\Psi,...)
        \,d(\tilde{R^\alpha}) \nonumber \\ &\simeq& \sum_{i=2}^{j}
        f(\bar{\Psi}_i,(\mathcal{D}\Psi)_i,...)
       (R^\alpha_i - R^\alpha_{i-1}),
\end{eqnarray}
where $f(...)$ is a placeholder for the right-hand sides of
(\ref{eq:b-from-chi-integated-by-parts}) and the integrated versions
of Eqs.~(\ref{Dlng}-\ref{DR32Psiu}), and we use $\alpha=9/2,2,2,3/2$, respectively, in these equations. We use this
discretisation of the integration measure because of its lower error
near the origin compared with $dR$.

Because our finite-differencing scheme is second-order accurate in
$\Delta x$, we expect any output to also converge to second order at
sufficiently early time. We have checked convergence with a sequence
$N_k:=N_0 \cdot 2^k$ of resolutions with $N_0=100$ and
$k=0...4$. Denoting by $Z_k$ the output of the code for fixed initial
data and $N_k$ grid points, we expect the quantity $\Delta Z_k = 4^k
\cdot (Z_{k+1}-Z_k)$ to be approximately independent of $k$.

\begin{figure}[!ht]
\centering
\includegraphics[scale=0.6]{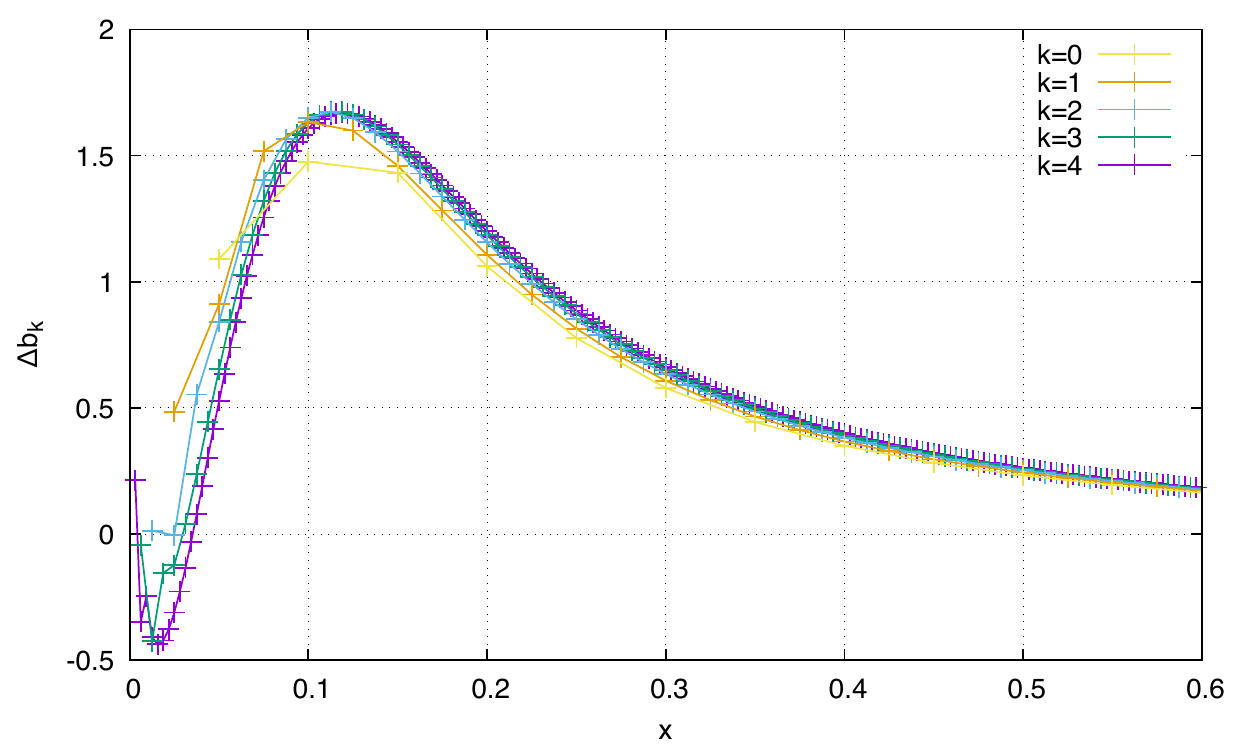}
\caption{The scaled error $\Delta b_k$ for $k=0...4$, represented for
$0 \leq x \leq 0.6$ and at a particular time instance $u=0.444$ for centered Gaussian pure gravitational wave initial data with $b(0,x)=13.88\exp[-(x/0.25)^2]$. While the curves progressively coincide for $x \gtrsim 0.2$, they differ slightly
at the first grid points, although some (slower) convergence is still
noticeable.}
\label{fig:convergence-testing}
\end{figure}

We found pointwise convergence to second order in $\Delta x$ in the
bulk of the grid, except near the origin. The error at the first
gridpoint was found to be approximately first-order. We have not found
a stable way of improving on this. The transition to second order is
illustrated in Fig.~\ref{fig:convergence-testing}.


The computation of the function $\Gamma(x)$, which appears in
Eq.~\eqref{eq:chi-evol-three-terms}, is done by performing a Taylor
expansion up to 7th order once its argument satisfies $bR^2 \leq
0.01$. This way, its zeroth and first order terms are manifestly
cancelled, avoiding numerical error near the origin from using the
full expression for $\Gamma$.

We diagnose the formation of a marginally outer-trapped surface by
comparing the maximum over one moment of time $u$ of the
compactness $\mathcal{C}:=M/R^2$, defined by Eq.~\eqref{compactness},
to a fixed threshold $\mathcal{C}_\text{max}=0.999$. Similarly, we
diagnose dispersion if the maximum of the compactness over the slice
of constant $u$ becomes smaller than $\mathcal{C}_\text{min}=0.001$.


\section{Similarity coordinates}
\label{section:similaritysolutions}


In any coordinates $x^\mu:=(T,\xi,\theta,\varphi, \psi)$ adapted to
the Bianchi symmetry and to DSS, by definition a spacetime is DSS if
and only if the metric takes the form
\begin{equation}
\label{eq:DSSmetric}
g_{\mu\nu}=e^{-2T}\tilde g_{\mu\nu},
\end{equation}
where $\tilde g_{\mu\nu}$ is periodic in $T$ with some period
$\Delta$. In particular, the area radius $R$ must take the form
\begin{equation}
\label{eq:hatRdef}
R=e^{-T}\hat R,
\end{equation}
with $\hat R$ again periodic.  A scalar-field $\Psi$ whose
stress-energy tensor is compatible with this metric must itself be
periodic in $T$ with the same period.

We now introduce the specific DSS-adapted coordinates
\begin{align}
\label{Tdef}
	T &:= - \ln\left( \frac{u_*-u}{k} \right)\\
	\xi &:= \frac{R}{u_*-u} = \frac{R}{k}e^{-T}
\end{align}
for a constant $u_*>0$ and $u<u_*$. (For $u>u_*$, both $\xi$ and $T$
are undefined). The constant $k$ is a length scale which we set to
$1$. From Eq.~\eqref{eq:uxmetric} it is clear that the metric in
coordinates $(\xi,T)$ is of the form \eqref{eq:DSSmetric}, and that the
spacetime is DSS if and only if $g$, $\hat R$ and
$B$ are periodic in $T$.

When either $\Psi(0,x)=0$ or $B(0,x)=0$, we expect all dimensionless
physical quantities, such as $\Psi$ or $B$ and $M/R^2$, to be periodic
in $T$ while the spacetime approximates the critical solution. We also
expect dimensionful quantities to scale as $e^{-lT}$, where $l$ is
their length dimension. Thus, in the pure scalar field critical
solution, $R_\Psi$ behaves as $e^{2T}$ times a periodic function of
$T$ (at constant $x$), and in the pure gravitational wave
critical solution, $R_B$ is $e^{2T}$ times a periodic function of
$T$.


\section{Numerical results}
\label{section:results}


\subsection{Initial data}


We choose the 2-parameter family of Gaussian initial data (with
parameters $p$ and $q$)
\begin{align}
    \Psi(0, x) &= p(1-q) A_{(\Psi)} \nonumber \\ & \exp \bigg[ -\bigg(
      \frac{R-\mu_{(\Psi)}}{w_{(\Psi)}} \bigg)^2
      \bigg], \label{Psi_initial_data} \\ 
\chi(0, x) &= pq A_{(\chi)}
    \exp \bigg[ -\bigg( \frac{R-\mu_{(\chi)}}{w_{(\chi)}} \bigg)^2
      \bigg], \label{b_initial_data}
\end{align}
as well as a two-parameter family with the profile of the derivative
of a Gaussian function:
\begin{align}
    \Psi(0, x) &= -2p(1-q) A_{(\Psi)} \left(\frac{R-\mu_{(\Psi)}}{w_{(\Psi)}^2}\right) e^{-\left( \frac{R-\mu_{(\Psi)}}{w_{(\Psi)}} \right)^2}, \label{Psi_initial_data-deriv} \\
    \chi(0, x) &= -2pq A_{(\chi)}  
    \left(\frac{R-\mu_{(\chi)}}{w_{(\chi)}^2}\right)
    e^{-\left( \frac{R-\mu_{(\chi)}}{w_{(\chi)}^2} \right)^2}. \label{b_initial_data-deriv}
\end{align}
Here $p q A_{(\Psi)}$ and $p (1-q) A_{(\chi)}$ are the amplitudes,
$w_{(\Psi)}$ and $w_{(\chi)}$ are the widths, and $\mu_{(\chi)}$ and
$\mu_{(\Psi)}$ the centres of the Gaussians. The free initial data
  for the evolved variables are completed by Eq.~(\ref{Rinitial}) above.

The field equations, with the gauge boundary condition $g=1$ at the
centre, are scale-invariant in the sense that they do not change when
we replace the arguments $(u,x)$ of $G$ (or $g$), $R$, $B$ (or $b$)
and $\Psi$ by $(\lambda u,\lambda x)$, and the value of $R$ by
$\lambda R$ and of $b$ by $\lambda^{-2}b$, but leaving the values of
$G$ (or $g$), $B$ and $\Psi$ unchanged. Put simply, everything scales
according to its dimension, with $u$, $x$ and $R$ having dimension
length, $b$ having dimension (length)${}^{-2}$ and $B$, $G$, $g$ and
$\Psi$ being dimensionless. We fix this overall scale freedom by
always setting the outer boundary of the grid to $x_\text{max}=8$.

For a fixed value of $q$, we start the bisection in $p$ with a large
value of $x_0$ close to $x_\text{max}$, adjusting it manually and
restarting the procedure until all individual simulations retain good
spatial resolution throughout their evolution. This is done by keeping
track of the grid point index of the location of the apparent horizon
formed in the supercritical steps: if $x_0$ is too large, the horizon
is formed at small $x$ and the dynamics are not well resolved
spatially. If $x_0$ is too small and for sufficient fine-tuning, the
apparent horizon is formed outside the spatial grid.

After some experimentation, we choose widths, centers and amplitudes
\begin{equation}
    \begin{split}
       A_{(\chi)} &= 1.0, \\
       \mu_{(\chi)} &= 0.5,\\
       w_{(\chi)} &= 0.05,
    \end{split}
    \quad
    \begin{split}
        A_{(\Psi)} &= 0.01, \\
        \mu_{(\Psi)} &= 1.15325, \\
        w_{(\Psi)} &= 0.115325
    \end{split}
\end{equation}
for the Gaussian initial data and
\begin{equation}
    \begin{split}
       A_{(\chi)} &= 0.023, \\
       \mu_{(\chi)} &= 0.74,\\
       w_{(\chi)} &= 0.074,
    \end{split}
    \quad
    \begin{split}
        A_{(\Psi)} &= 0.034, \\
        \mu_{(\Psi)} &= 1.22, \\
        w_{(\Psi)} &= 0.224
    \end{split}
\end{equation}
for the Gaussian derivative initial data.
These have the following properties:

1. For pure scalar initial data $q=0$
  and pure gravitational wave initial data $q=1$, the critical
  amplitudes are $p\simeq 1$. This is essentially a matter of
  convenience.

2. For the two pure initial data sets the accumulation point of echos
at $R=0$, $u=u_*$, $v=v_*$ is at a similar value of $v_*$. This is
achieved in practice by independently finding the approximate value of
$x_0 \simeq v_*$ for two sets of initial data corresponding to pure
scalar field and pure gravitational waves, and then rescaling the
scalar field initial data such that the two values of $x_0$ coincide.

This ensures that when we choose values of $q$ representing a mixture
of the two fields and then fine-tune $p$ again to the threshold of
collapse, we can expect the fields to interact strongly. By contrast,
if $v_*$ was much smaller for, say, the scalar field, in fine-tuning
$p$ for mixed data to the threshold of collapse, we would be likely to
find critical collapse dominated by the scalar field, with the
gravitational waves arriving later and either dispersing or forming a
large black hole.

All plots in the following correspond to the Gaussian initial data,
except for Fig.~\ref{fig:gamma-q}, which compares results from the two
families.

For given $q$, we perform 50 bisection steps from a rough initial
bracket for $p_*(q)$ to determine its value up to machine
precision. We work in double precision. With $p_*(q)$ known (for a
given set of numerical parameters such as $x_0$, $x_\text{max}$, $C$
and $\Delta x$) the scaling laws are then re-evaluated on 450 evenly
spaced points in $\log_{10}|p-p_*|$, with 30 points per decade, to
resolve for the fine structure of the DSS scaling, which we expect to
be periodic with period $\Delta/(\ln(10)\gamma)$ in
$\log_{10}|p-p_*|$.

\subsection{The pure field cases}

The mass and curvature scaling laws obtained for pure scalar field
($q=0$) and pure gravitational wave ($q=1$) initial data give critical
exponents $\gamma_{\Psi} \simeq 0.415$ and $\gamma_{B} \simeq 0.164$
respectively, which agree with the results found in 
  \cite{Bland_Preston_Becker_Kunstatter_Husain_2005} and
  \cite{Bizon_Chmaj_Schmidt_2005}, respectively.

The echoing periods $\Delta$ [in $T$, defined above in
    Eq.~(\ref{Tdef})] of the best near-critical solutions were
estimated by identifying the period with that of the Fourier mode of
highest peak of $\Psi$ or $B$, and then fitting the curves by eye with
a sine wave of the same period. We determined $\Delta_{\Psi} \simeq
1.6$ and $\Delta_B \simeq 0.47$, in agreement with the values
  found in \cite{Bland_Preston_Becker_Kunstatter_Husain_2005} and
  \cite{Bizon_Chmaj_Schmidt_2005}.


\subsection{Gravitational waves with small scalar field perturbation}


We now add a small perturbation $\varepsilon \ll 1$ to both $q=0$ and
$q=1$, so that either $b$ or $\Psi$ evolves as an almost-linear
  perturbation on a background solution driven by the other
  field. 

We begin with the case $q=1-\varepsilon$, with
$\varepsilon=10^{-6}$. When $\Psi$ evolves essentially as a linear
perturbation, separation of variables allows us to consistently look
for solutions of the scalar test field equation of the form
\begin{equation}
\label{eq:psi-as-a-perturbation}
    \Psi(\xi, T) = {\rm Re}\,e^{\lambda_\Psi T} \hat{\Psi} (\xi, T),
\end{equation}
where $\lambda_\Psi = \kappa_\Psi+i \omega_\Psi$ is a complex number
and the complex function $\hat{\Psi} (\xi, T)$ is periodic in
$T$ with period $\Delta_B$ (the same as the background
solution). As a result, $e^{-\kappa_\Psi T}\Psi(\xi, T)$ is
only quasi-periodic in $T$, with a discrete spectrum offset by
$\omega_\Psi$.

\begin{figure}[!ht]
\centering
\includegraphics[scale=0.62]{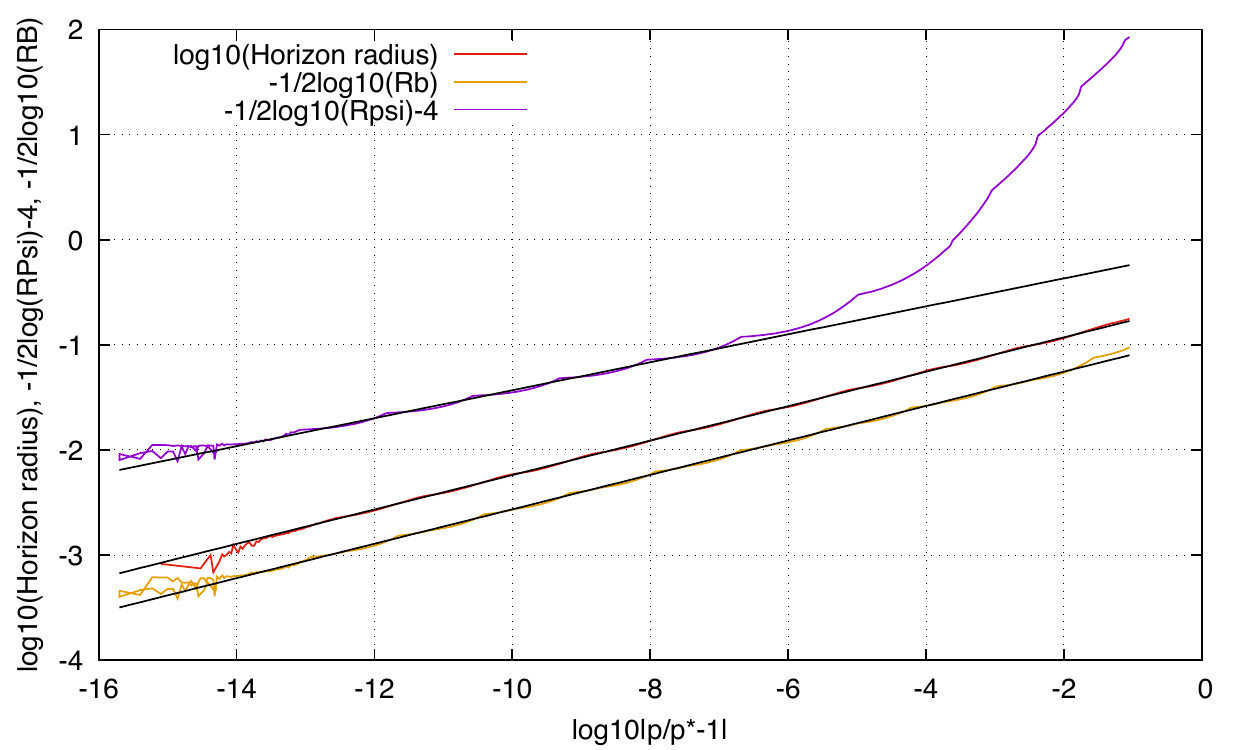}
\caption{Scaling laws for the radius $R_\text{ah}$ of apparent horizon
  formation, and the global maximum of the Ricci scalar $R_\Psi$ and
  of $R_B$ for the case $q=1-\varepsilon$ and Gaussian initial
  data. These two last quantities are rescaled by $-1/2$ in the log
  plot to account for their dimension $\text{length}^{-2}$. The black lines
  represent the linear fits to each curve. The slope of the lines
  fitted against $R_\text{ah}$ and $R_B^{-1/2}$ are $0.1638$, and
  $0.133$ for $R_\Psi^{-1/2}$.}
\label{fig:1eps-scaling}
\end{figure}

The radius $R_\text{ah}$ of apparent horizon formation, which has
dimension length, scales as
\begin{equation}
R_\text{ah}(p)\sim (p-p_*)^{\gamma_B}.
\end{equation}
By applying \eqref{eq:psi-as-a-perturbation} to the expression for
$R_\Psi$, which has dimension $\text{length}^{-2}$, we deduce that it
scales as $\sim e^{2(1+\kappa_\Psi)T}$ when the scalar field is
treated perturbatively. For near-critical solutions, the maximum value
of curvature is achieved just after departing from
self-similarity, which occurs at a time $T \simeq -\gamma_B \ln |p -
p^*|$ \cite{Hod_Piran_1997}. From this we obtain the scaling relation
\begin{equation}
\label{eq:Ricci-scaling-as-perturbation}
    \left(\max_{\xi,T} R_\Psi\right)^{-1/2} \sim (p-p_*)^{(1+\kappa_\Psi)\gamma_B}.
\end{equation}
The critical exponents $\gamma_B \simeq 0.164$ and
$\tilde{\gamma}_{\Psi}=(1+\kappa_\Psi)\gamma_B \simeq 0.133$
were calculated from the mass and curvature scaling laws for
$q=1-\varepsilon=1-10^{-6}$ (Fig.~\ref{fig:1eps-scaling}), giving us
$\kappa_\Psi \simeq -0.19$.

\begin{figure}[!ht]
\centering
\includegraphics[scale=0.7]{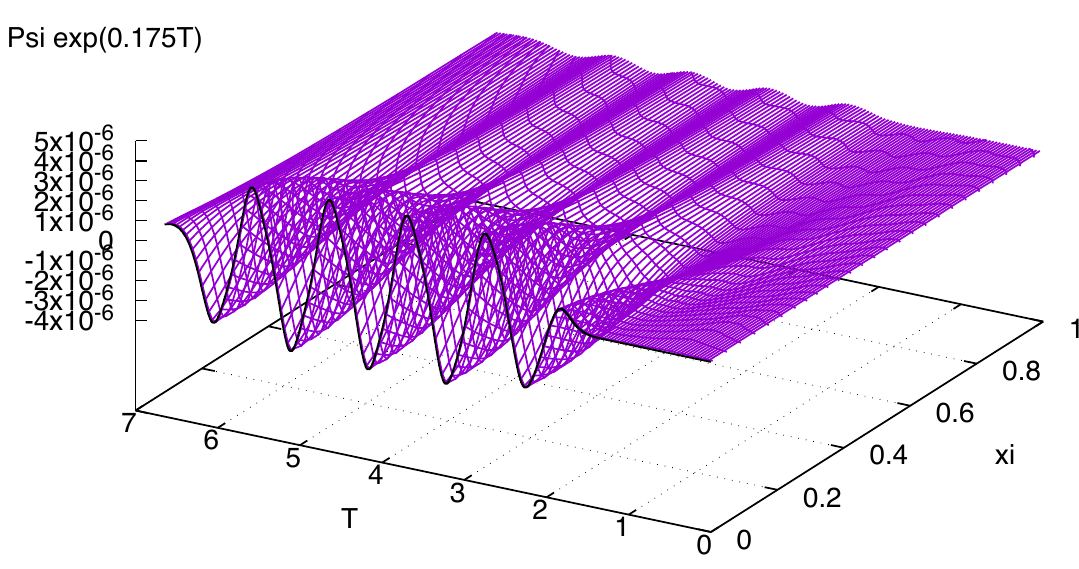}
\caption{The scalar field $\Psi(\xi, T) e^{-\kappa_\Psi T}$ for
  optimal fine-tuning with $q=1-\varepsilon$, $\kappa_\Psi =
  -0.175$. A black line represents the extrapolation to the regular
  centre $R=0$. }
\label{fig:3d-1eps-psi}
\end{figure}

The perturbation exponent $\kappa_\Psi$ was independently estimated by
adjusting $\Psi e^{-\kappa_\Psi T}$ by eye to be as quasi-periodic as
possible in our best near-critical evolution, placing it in the
interval $\kappa_\Psi \in (-0.2, -0.15)$ (see
Fig.~\ref{fig:3d-1eps-psi}).

\begin{figure}[!ht]
\centering
\includegraphics[scale=0.62]{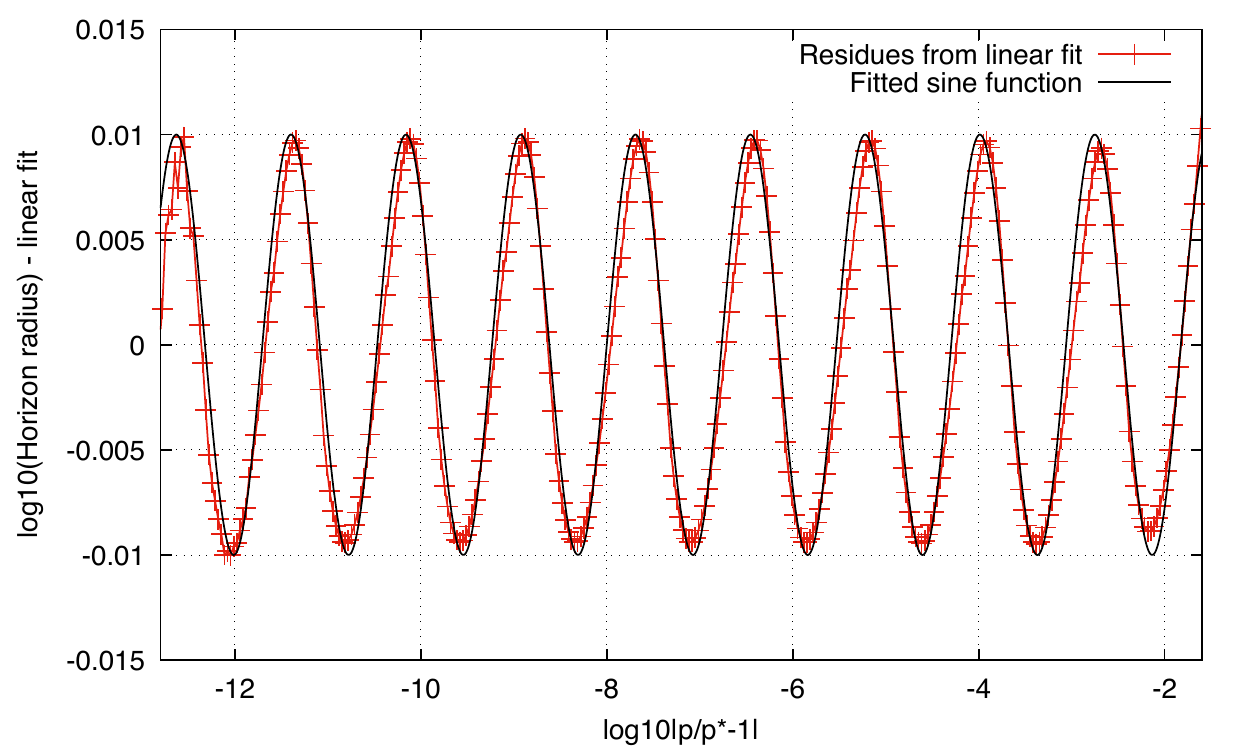}
\caption{Residuals of the linear fit to
    Fig.~\ref{fig:1eps-scaling} for the radius $R_\text{ah}$ of
  apparent horizon formation for $q=1-\varepsilon$. The scaling
  exponent is $\gamma = 0.1638$ and the fitted period of the residuals
  is $\Delta_\text{res} = 1.235$, which is related to the echoing
  period of the critical solution by $\Delta_\text{res} =
  \Delta_B/(\ln(10)\gamma)$, resulting in $\Delta_B \simeq 0.47$,
  consistent with Fig.~\ref{fig:3d-1eps-compactness}.}
\label{fig:1eps-hr-residuals}
\end{figure}

\begin{figure}[!ht]
\centering
\includegraphics[scale=0.62]{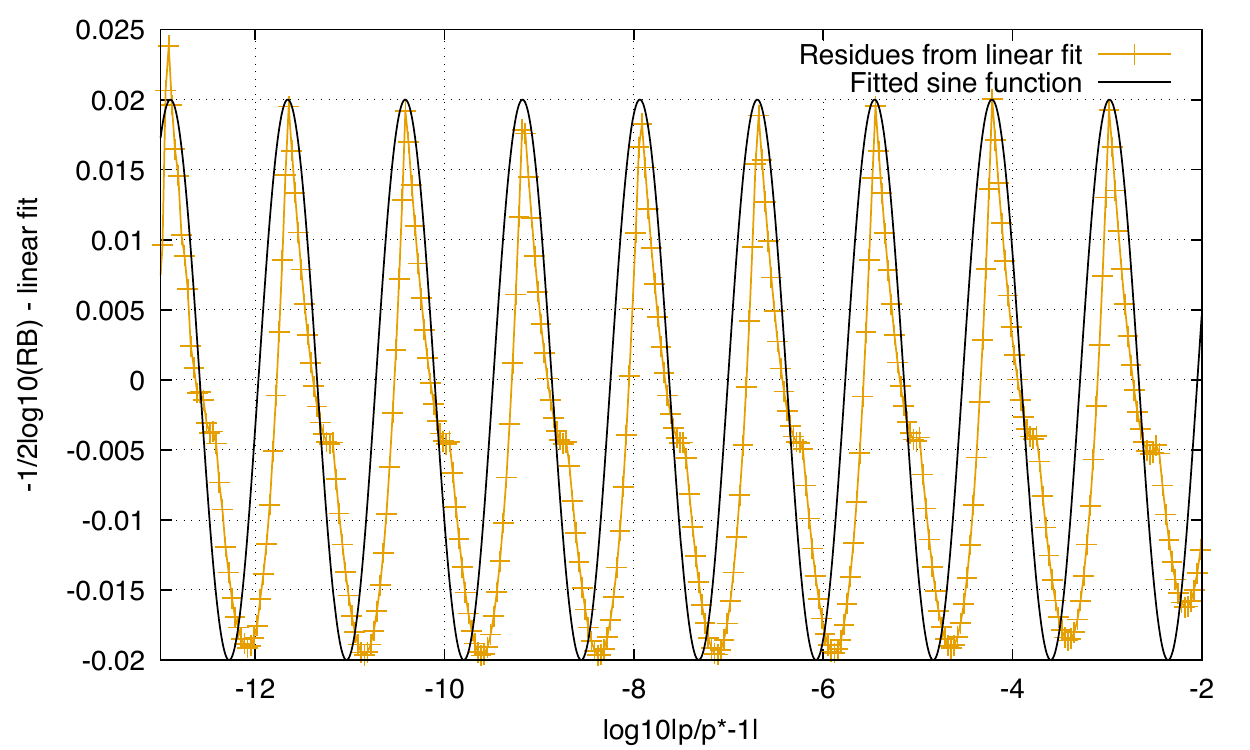}
\caption{Residuals of the linear fit to
    Fig.~\ref{fig:1eps-scaling} for $R_B$ for
  $q=1-\varepsilon$. The scaling exponent is $\gamma = 0.1638$ and the
  fitted period of the residuals is $\Delta_\text{res} = 1.24$,
  resulting in $\Delta_B \simeq 0.47$, consistent with
  Fig.~\ref{fig:3d-1eps-B}.}
\label{fig:1eps-rb-residuals}
\end{figure}

\begin{figure}[!ht]
\centering
\includegraphics[scale=0.62]{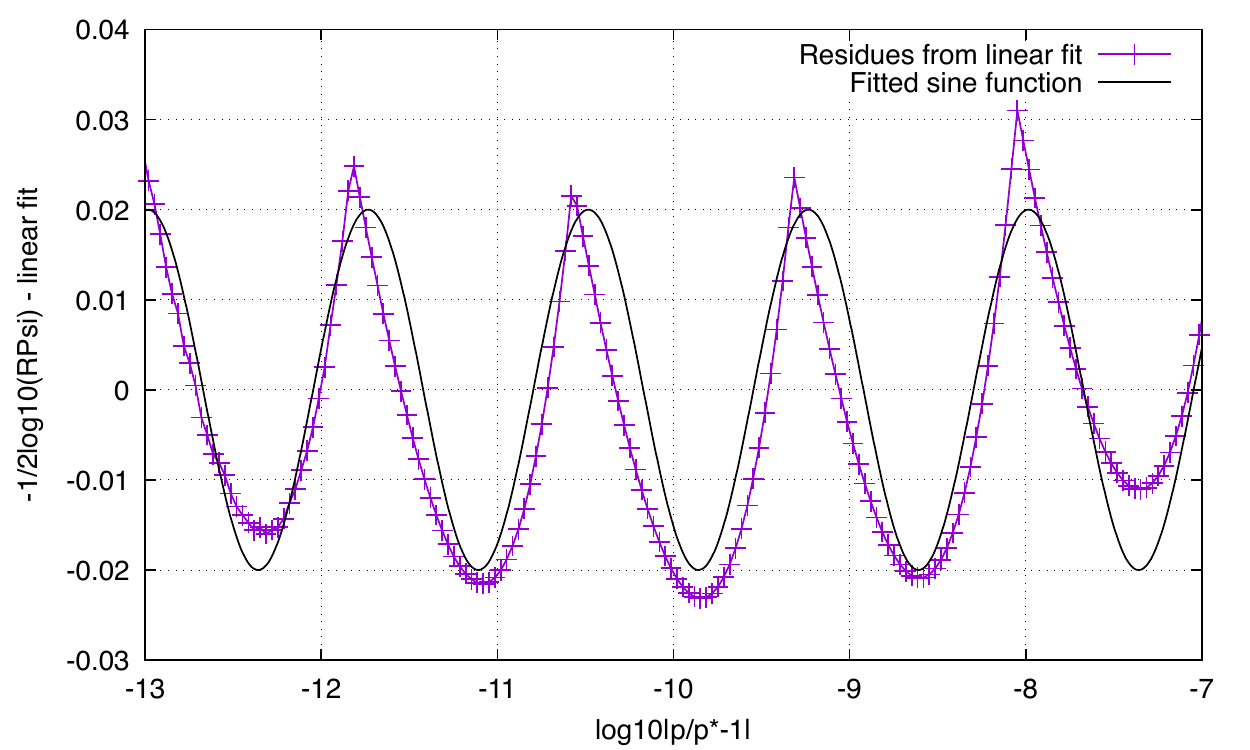}
\caption{Residuals of the linear fit to
    Fig.~\ref{fig:1eps-scaling} for the Ricci scalar $R_\Psi$ for $q=1-\varepsilon$. The scaling exponent is $\gamma = 0.133$ and the fitted period of the residuals is $\Delta_\text{res} = 1.25$.}
\label{fig:1eps-rpsi-residuals}
\end{figure}

Fig.~\ref{fig:1eps-hr-residuals} shows the residuals of the
linear fit of the scaling law for $R_\text{ah}$
\begin{equation}
\text{res}(p):=\log_{10} R_\text{ah}-\gamma \log_{10}|p-p_*| - \beta,
\end{equation}
where $\beta$ is the intercept of the fit. Similar plots for
$R_B^{-1/2}$ and $R_\Psi^{-1/2}$ are shown in
Figs.~\ref{fig:1eps-rb-residuals}-\ref{fig:1eps-rpsi-residuals}.

\begin{figure}[!ht]
\centering
\includegraphics[scale=0.75]{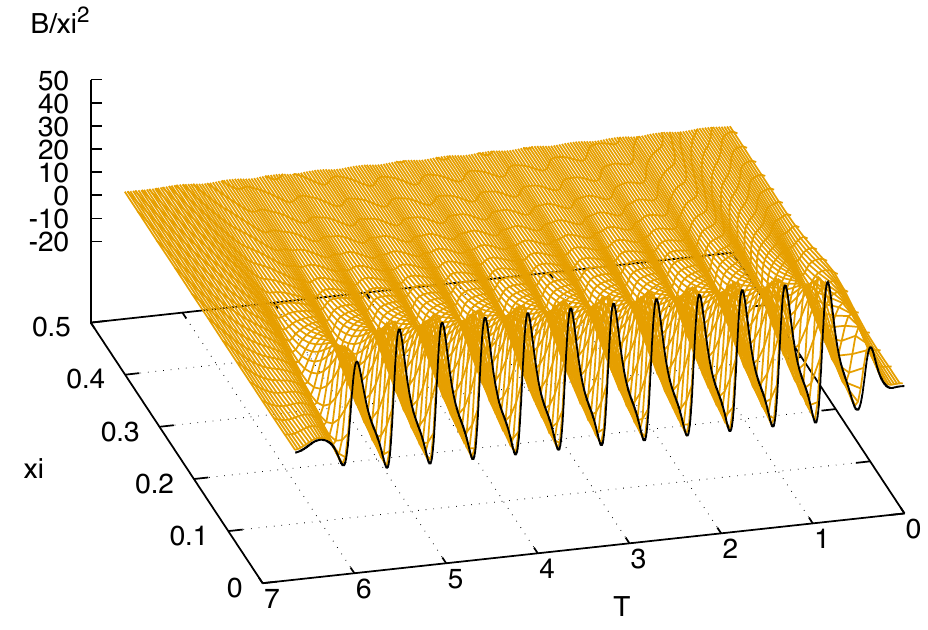}
\caption{The quantity $B/\xi^2$ for optimal fine-tuning with $q=1-\varepsilon$. A black line represents the extrapolation to the regular centre $R=0$. }
\label{fig:3d-1eps-B}
\end{figure}

\begin{figure}[!ht]
\centering
\includegraphics[scale=0.8]{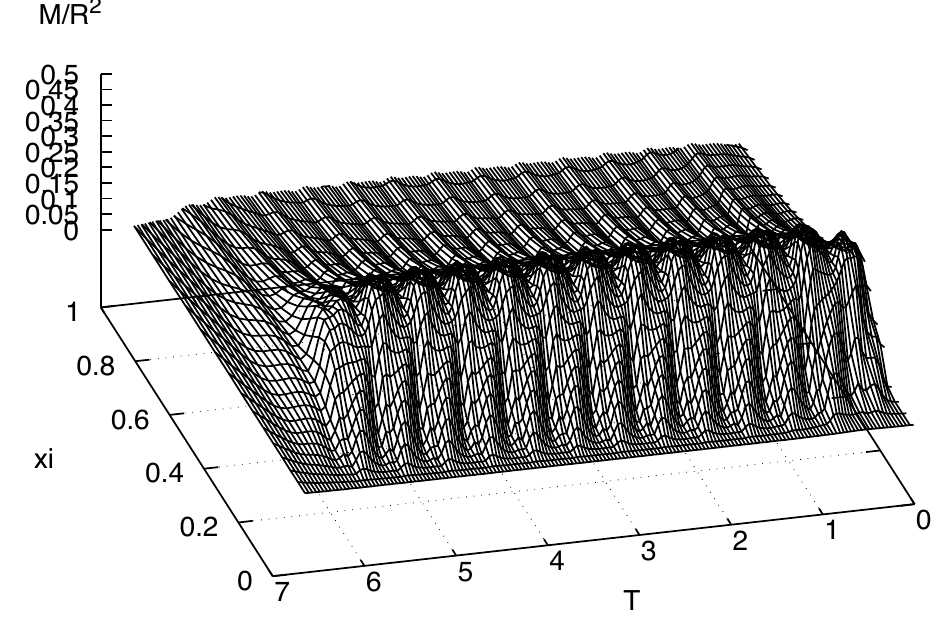}
\caption{The compactness $M/R^2$ for optimal fine-tuning with $q=1-\varepsilon$. }
\label{fig:3d-1eps-compactness}
\end{figure}

The quantities $\Psi$, $B/\xi^2$ and $M/R^2$ are represented in
Figs.~\ref{fig:3d-1eps-psi}-\ref{fig:3d-1eps-compactness} for the
best subcritical evolution for Gaussian initial data.  Both $B$ and
$B/\xi^2$ are dimensionless, but $B$ is $O(R^2)$ near the origin while
$B/\xi^2$ is $O(1)$, which is why we plot the latter.  Note that
  because $\epsilon$ is small, at this resolution
  Figs.~\ref{fig:1eps-hr-residuals}, \ref{fig:1eps-rb-residuals},
  \ref{fig:3d-1eps-B} and \ref{fig:3d-1eps-compactness} are
  indistinguishable from their counterparts in the pure gravitational
  wave case $q=1$, so they can serve to illustrate that case, too.


\subsection{Scalar field with small gravitational wave perturbation}

\begin{figure}[!ht]
\centering
\includegraphics[scale=0.62]{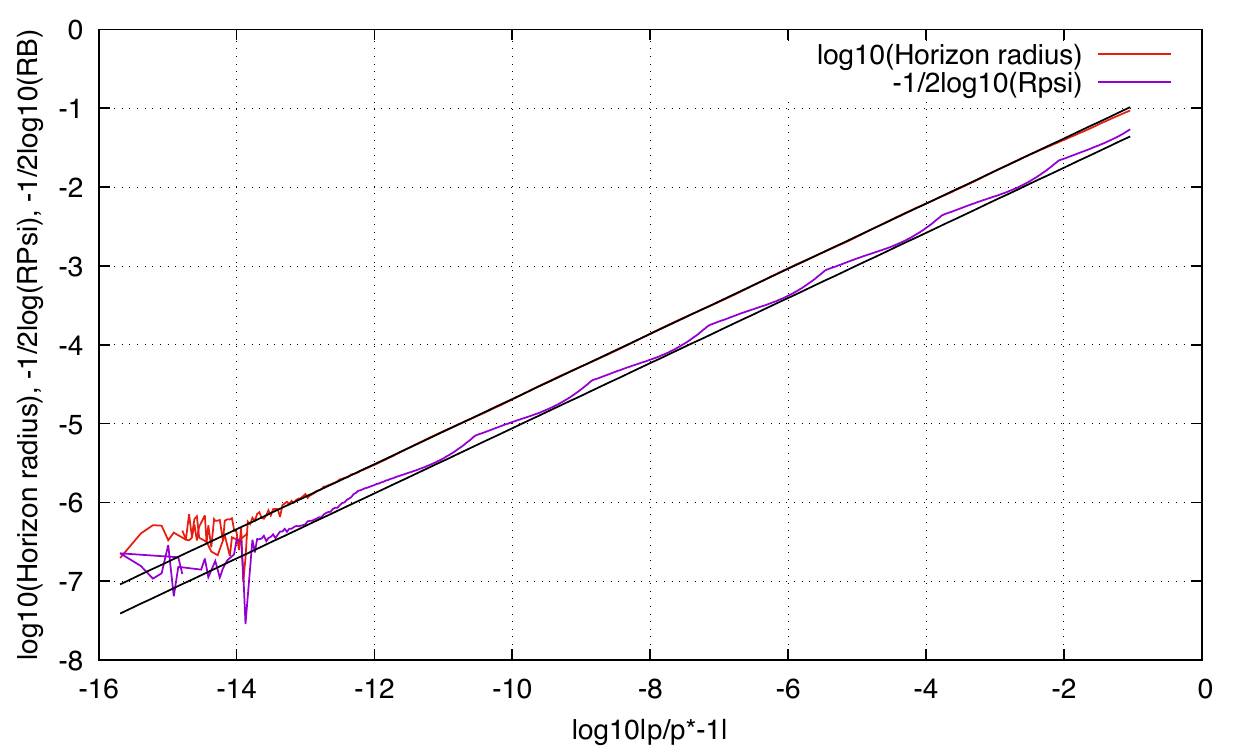}
\caption{Scaling laws for the radius $R_\text{ah}$ of apparent horizon
  formation, and the global maximum of the Ricci scalar $R_\Psi$ for
  the case $q=\varepsilon$ and Gaussian initial data. The latter is
  rescaled by $-1/2$ in the log plot to account for its dimension
  $\text{length}^{-2}$. The black lines represent the linear fits to
  each curve. The slope of the lines fitted against $R_\text{ah}$ and
  $(R_\Psi)^{-1/2}$ were $0.4131$.}
\label{fig:0eps-scaling}
\end{figure}

Similar calculations hold for $B$ and $R_B$ when
$q=\varepsilon=10^{-6}$, where the gravitational waves are treated as
a linear perturbation on the dominant scalar field solution, giving us
\begin{equation}
        B(\xi, T) =  {\rm Re}\, e^{\lambda_B T} \hat{B} (\xi, T),
\end{equation}
with $\lambda_B = \kappa_B + i \omega_B$. The critical exponent
$\gamma_\Psi \simeq 0.413$ was calculated numerically from the scaling
laws for the radius of apparent horizon formation and for the Ricci
scalar, see Fig.~(\ref{fig:0eps-scaling}). The perturbation exponent
$\kappa_B$ was estimated by adjusting $B e^{-\kappa_B T}$ by eye to be
as periodic as possible in our best near-critical evolution, placing
it in the interval $\kappa_B \in (-1.55, -1.45)$. The maximum of
the pseudo-curvature $R_B$ does not show power law scaling in $(p-p_*)$:
$R_B$ scales as $\sim e^{2(1+\kappa_B)T}$, which decays because
$\kappa_B < -1$, and so its global maximum is dominated by a value at
early times which is dependent on the initial data, and so one cannot
apply the same argument that led to
Eq.~\eqref{eq:Ricci-scaling-as-perturbation}.

\begin{figure}[!ht]
\centering
\includegraphics[scale=0.62]{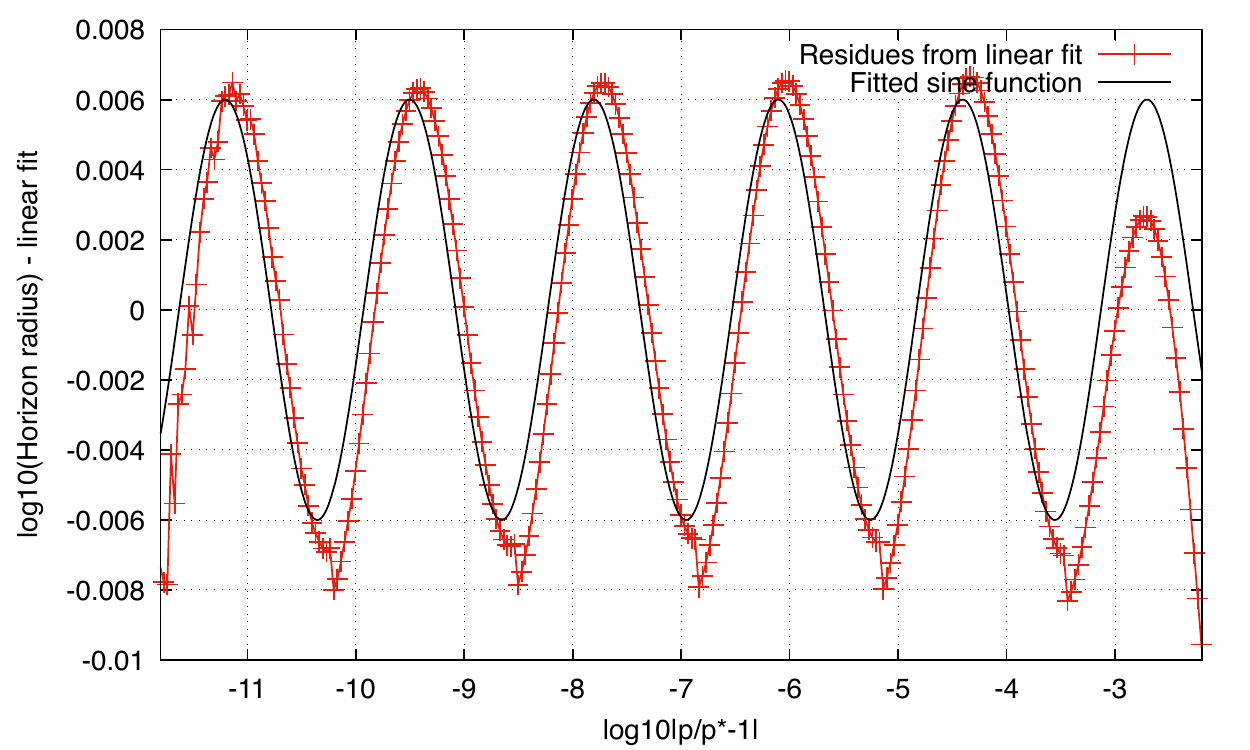}
\caption{Residuals of the linear fit to
    Fig.~\ref{fig:0eps-scaling} for the for the radius
  $R_\text{ah}$ of apparent horizon formation for $q=\varepsilon$. The
  scaling exponent is $\gamma = 0.4131$ and the fitted period of the
  residuals is $\Delta_\text{res} = 1.7$, which is related to the
  echoing period of the critical solution by $\Delta_\text{res} =
  \Delta_\Psi/(\ln(10)\gamma)$, resulting in $\Delta_\Psi \simeq 1.6$,
  consistent with Fig.~\ref{fig:3d-0eps-compactness}.}
\label{fig:0eps-hr-residuals}
\end{figure}

\begin{figure}[!ht]
\centering
\includegraphics[scale=0.62]{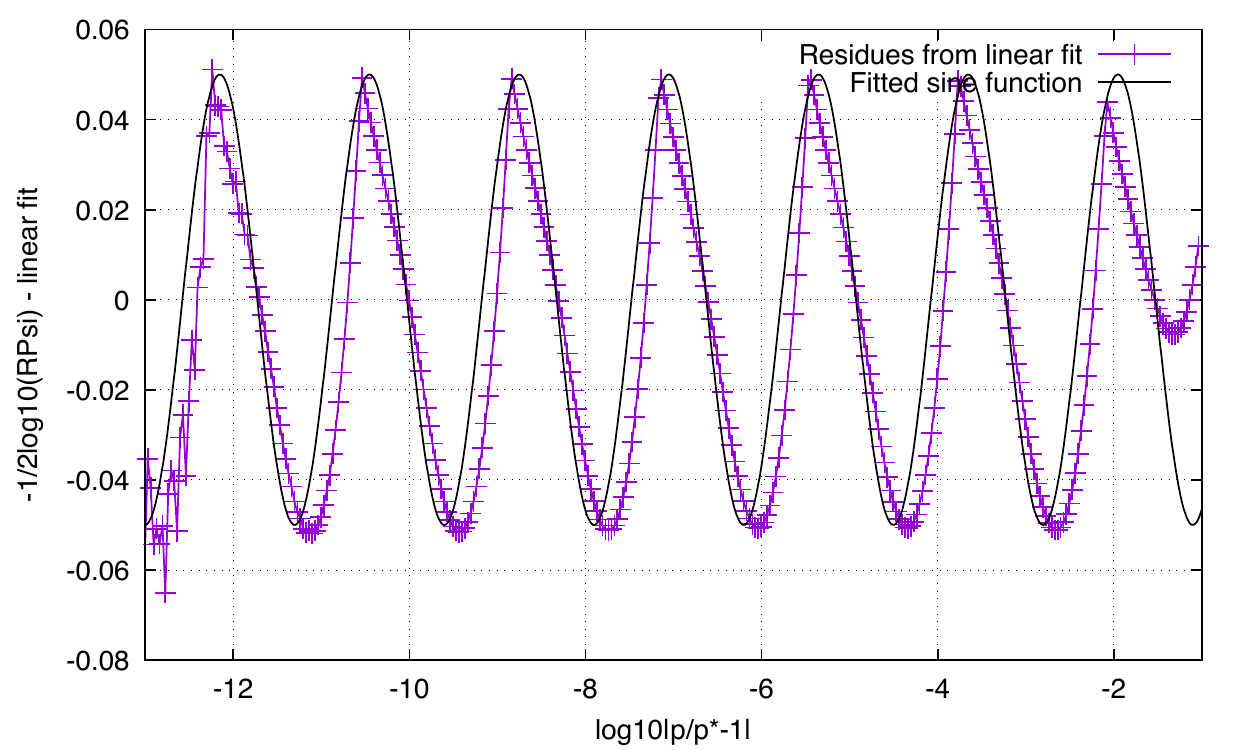}
\caption{Residuals of the linear fit to
    Fig.~\ref{fig:0eps-scaling} for the Ricci scalar $R_\Psi$ for
  $q=\varepsilon$. The scaling exponent is $\gamma = 0.4131$ and the
  fitted period of the residuals is $\Delta_\text{res} = 1.7$, which
  is related to the echoing period of the critical solution by
  $\Delta_\text{res} = \Delta_\Psi/(\ln(10)\gamma)$, resulting in
  $\Delta_\Psi \simeq 1.6$, consistent with
  Fig.~\ref{fig:3d-0eps-psi}.}
\label{fig:0eps-rpsi-residuals}
\end{figure}

\begin{figure}[!ht]
\centering
\includegraphics[scale=0.7]{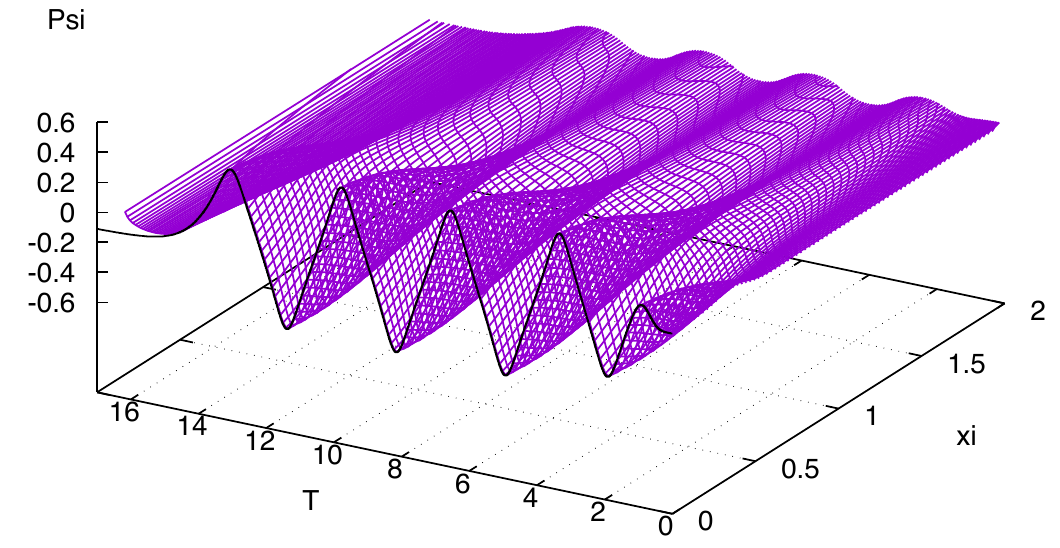}
\caption{The scalar field $\Psi(\xi, T)$ for optimal fine-tuning with
  $q=\varepsilon$. A black line represents the extrapolation to the
  regular centre $R=0$. }
\label{fig:3d-0eps-psi}
\end{figure}

\begin{figure}[!ht]
\centering
\includegraphics[scale=0.8]{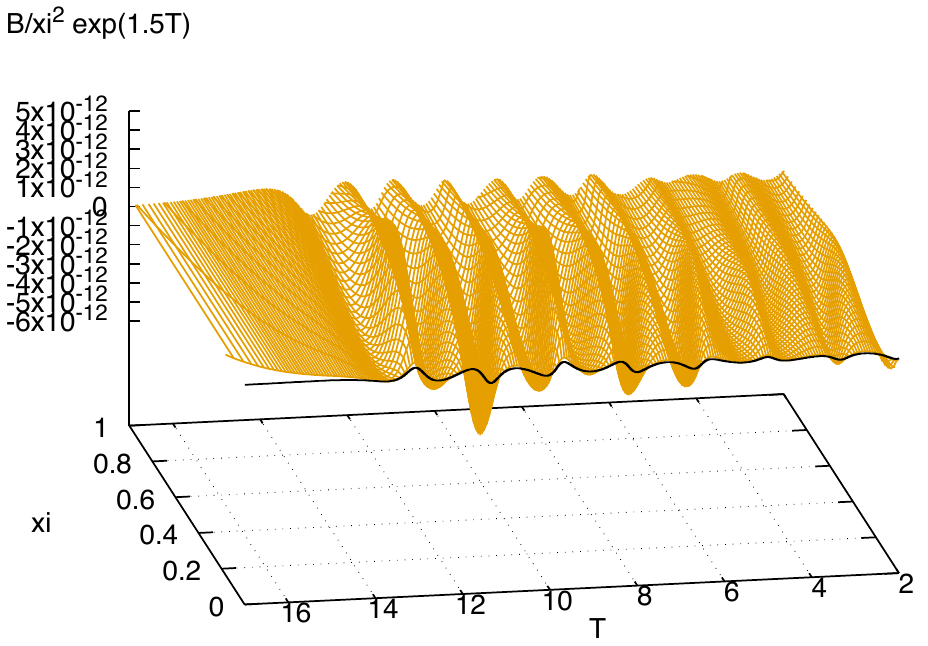}
\caption{The quantity $B/\xi^2 e^{-\kappa_B T}$ for optimal
  fine-tuning with $q=\varepsilon$, $\kappa_B = -1.5$. A black line
  represents the extrapolation to the regular centre $R=0$. $T$ is
  restricted to $[2, 17]$ so as to visualize the exponential
  correction to $B$ after the dominant scalar field starts to
  approximate the critical solution.}
\label{fig:3d-0eps-B}
\end{figure}

\begin{figure}[!ht]
\centering
\includegraphics[scale=0.8]{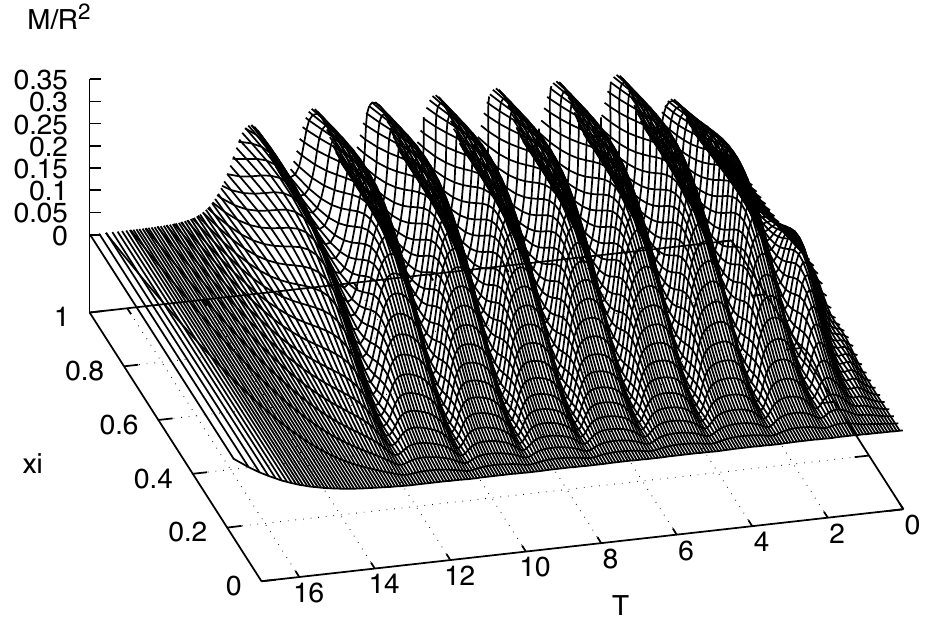}
\caption{The compactness $M/R^2$ for optimal fine-tuning with
  $q=\varepsilon$.}
\label{fig:3d-0eps-compactness}
\end{figure}

The residuals of the linear fit for the scaling laws of $R_\text{ah}$
and $R_\Psi^{-1/2}$ are represented in
Figs.~\ref{fig:0eps-hr-residuals}-\ref{fig:0eps-rpsi-residuals}. The
quantities $\Psi, B/\xi^2$ and $M/R^2$ are represented in
Figs.~\ref{fig:3d-0eps-psi}-\ref{fig:3d-0eps-compactness} for the best
subcritical evolution for Gaussian initial data.  Again,
Figs.~\ref{fig:0eps-hr-residuals}, \ref{fig:0eps-rpsi-residuals},
\ref{fig:3d-0eps-psi} and \ref{fig:3d-0eps-compactness} are at this
resolution indistinguishable from their counterparts in the case $q=0$
of a pure scalar field.


\subsection{Mixed fields and the bi-critical solution}


As $\kappa_B$ and $\kappa_\Psi$ have both negative real part, both
$\Psi$ and $B$ are decaying perturbations on the background critical
solution of the other field when their initial amplitude is
sufficiently small such that their dynamics are essentially linear.

When $q$ is decreased more from $q=1$, the scalar field $\Psi$ still
decays, but when $q \lesssim 0.9$ (for Gaussian initial data) the
non-linear dynamics play a more significant role and $\Psi$ instead
starts growing with $T$, with $R_\Psi$ eventually dominating $R_B$,
and the solution approaches the known scalar field critical solution
for large enough $T$. The same behaviour is observed for the other
2-parameter family of initial data, although the value of $q$ for
which the scalar field begins to grow with $T$ is $q \lesssim
0.85$. We have investigated the transition between these two regimes,
such that the scalar field and the gravitational wave both neither
grow nor decay in the critical solution found by fine-tuning $p$ to
$p_*$ for given $q\simeq q_*$. In other words, we have to fine-tune in
two parameters at once. In practice, we fine-tune to the black-hole
threshold $p=p_*(q)$ in an automated inner loop, and fine-tune to
$q_*$ in a manual outer loop, as the bisection criterion for $q$ is
less clear-cut than collapse versus dispersion for $p$, and we were
not sure what to expect at the $q$ threshold.

 We expect the bi-critical solution to be an intermediate
attractor for $(p,q)\simeq (p_*(q_*),q_*)$, in which the solution
becomes  at least approximately self-similar, with both fields
neither growing nor decaying.

In the triaxial vacuum collapse case investigated in
\cite{Bizon_Chmaj_Schmidt_2006}, for which the two competing fields
play symmetric roles ( the two critical solutions are the same
  up to a discrete symmetry), the bi-critical solution was also
found to be discretely self-similar with a constant echoing period. In
the present biaxial case plus scalar field, however, the two critical
solutions are distinct, with $\gamma_\Psi > \gamma_B$ and $\Delta_\Psi
> \Delta_B$.

    
\begin{figure*}
\centering 
\subfloat[$q:=q_a=0.9184570312$.  
\label{fig:mixed-q-low}]{
\includegraphics[width=0.5\textwidth]{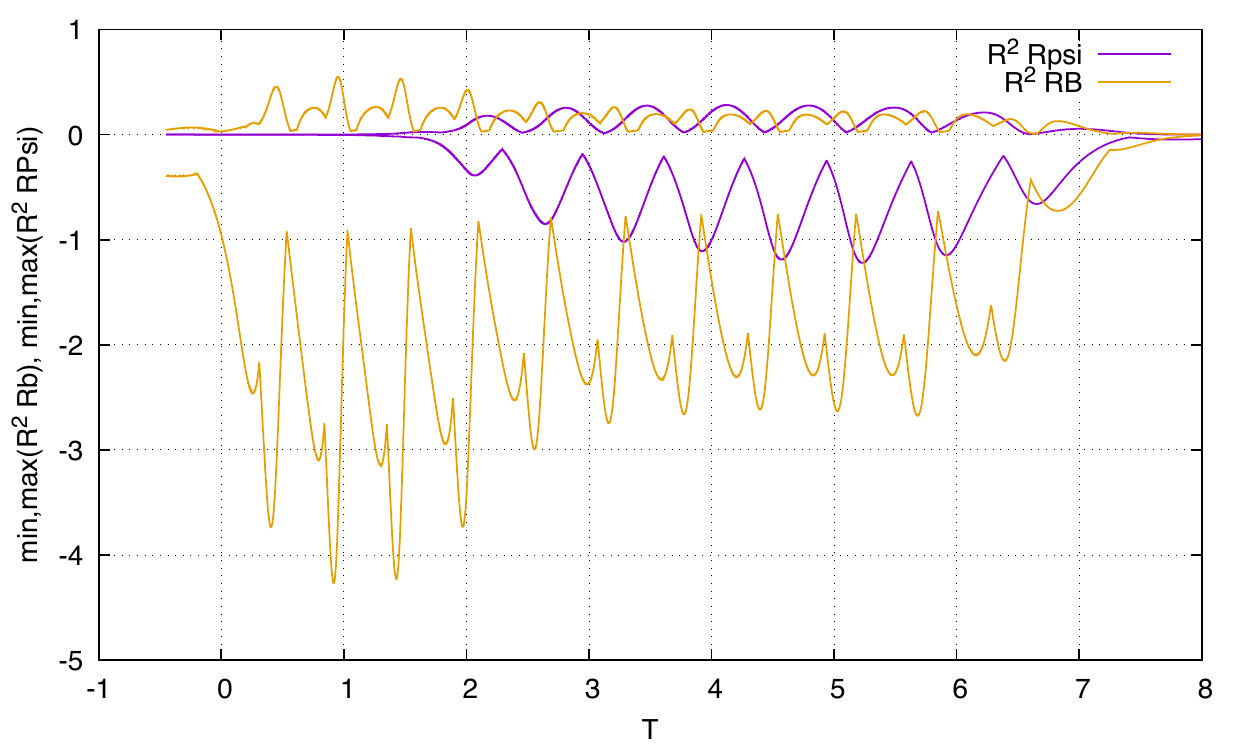}}
\subfloat[$q:=q_b=0.9200439452$.  
\label{fig:mixed-q-middle}]{
\includegraphics[width=0.5\textwidth]{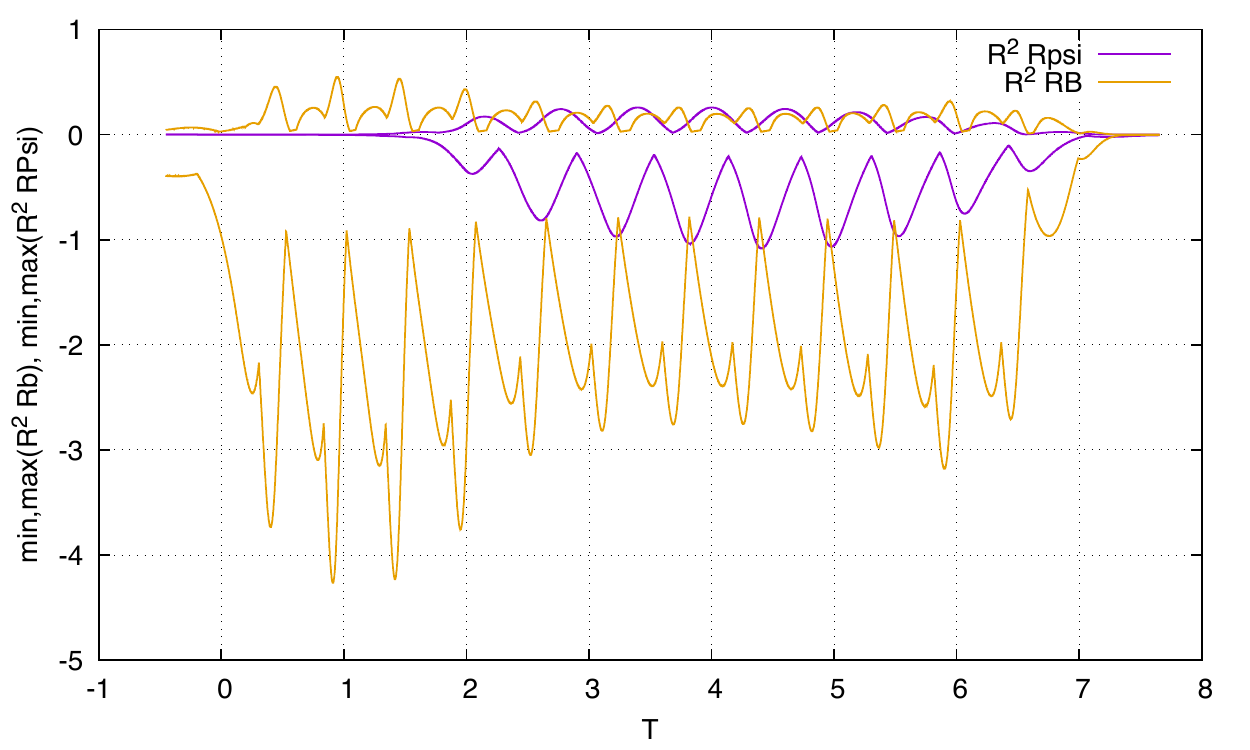}}
\newline
\subfloat[$q:=q_c=0.9216308593$.        
\label{fig:mixed-q-high}]{   
\includegraphics[width=0.5\textwidth]{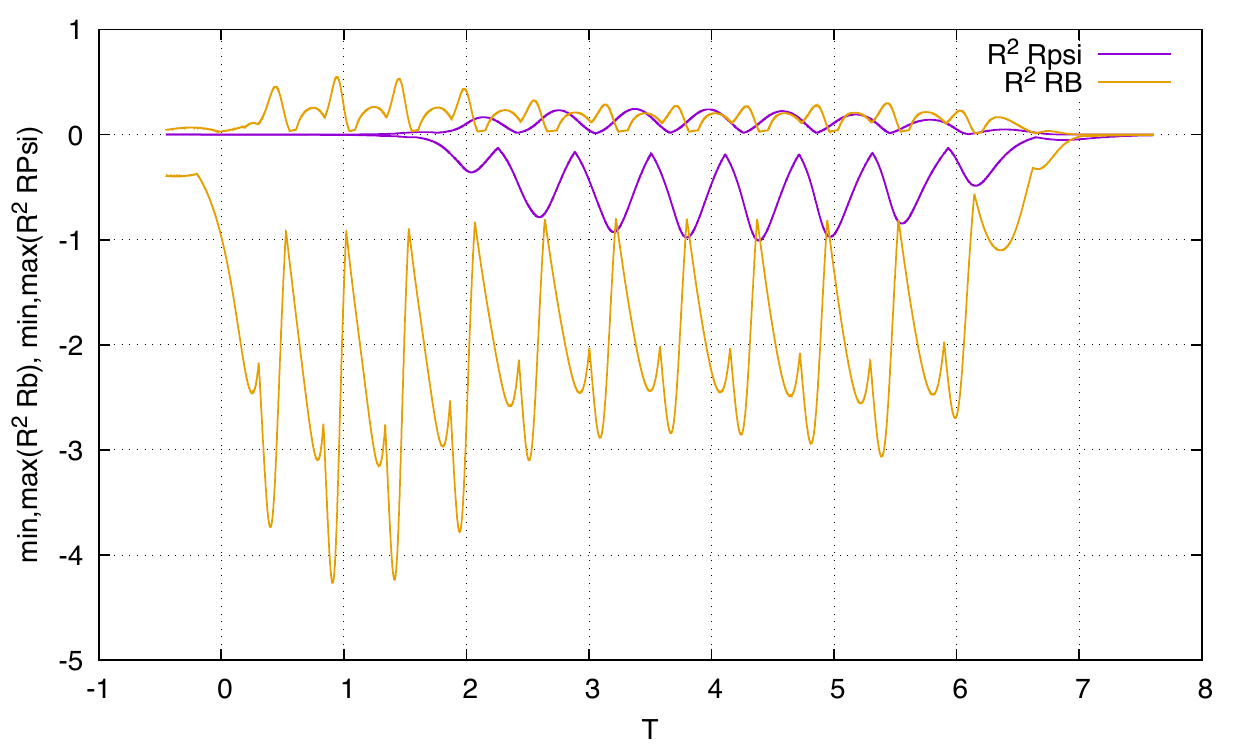}}
\subfloat[$q=0$ (purple) and $q=1$ (orange).        
\label{fig:mixed-q-pure-cases-references}]{   
\includegraphics[width=0.5\textwidth]{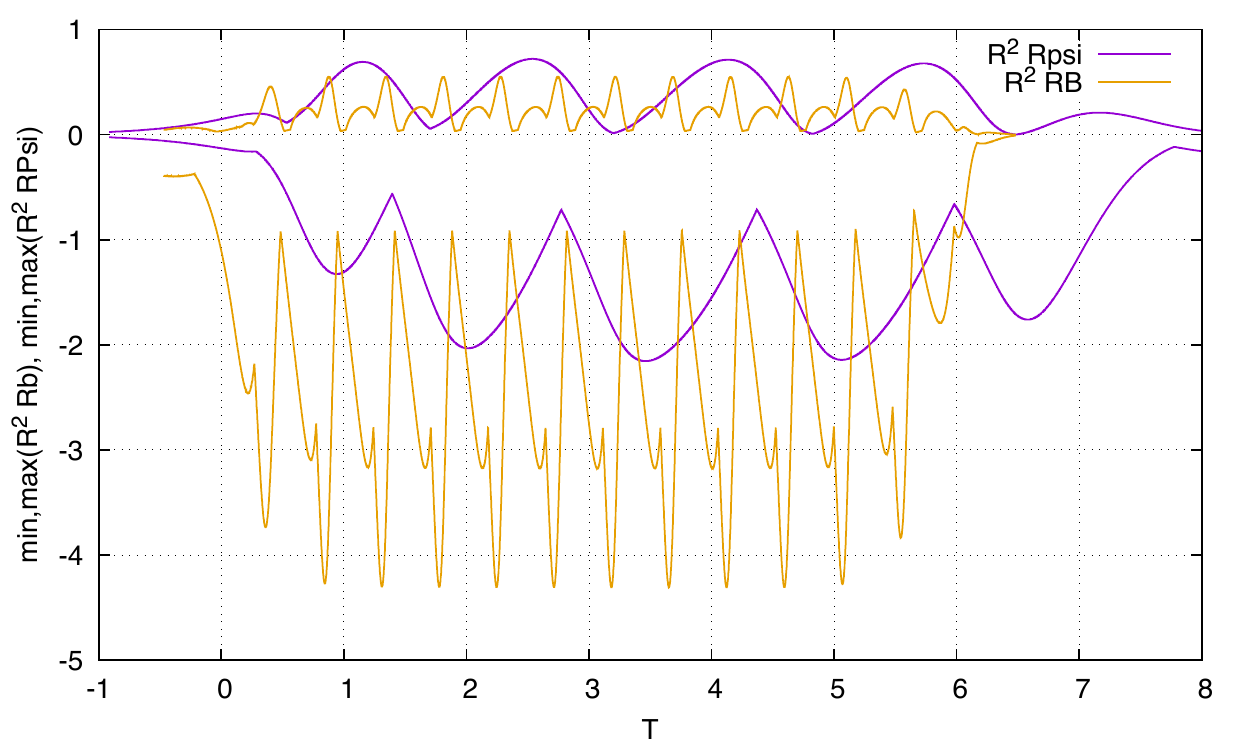}}
\caption{The maxima and minima (over $x$) of the quantities
  $R^2 R_B$ (orange) and $R^2 R_\Psi$ (purple), plotted against $T$
  for different values of $q$, with $q_a < q_b < q_c$, extracted from
  the respective best subcritical evolutions. For reference, the same
  quantities for the two pure critical solutions are plotted together in
  Fig.~\ref{fig:mixed-q-pure-cases-references}.}
\label{fig:mixed-q}
\end{figure*}
    

We would have expected that for $q\simeq q_*$ and $p$ sufficiently
close to $p_*(q)$, the solution starts out with both $b$ and $\Psi$
equally important. But this is not so at least for our two 2-parameter
families. Rather, in these solutions $\Psi$ starts out as a growing
perturbation of the $b$ critical solution, before entering a phase
where both $\Psi$ and $b$ neither grow nor decay, and spacetime is
still approximately DSS.

The presence of this transition phase means that we use up some of the
available fine-tuning of $p$, and hence some of the available range of
$T$, before we reach the expected bi-critical solution. This in turn
means that we cannot fine-tune $q$ as well as expected, nor observe
the properties of the bi-critical solution over as many periods as
expected.

Fig.~\ref{fig:mixed-q} illustrates the dimensionless quantities $R^2
R_\Psi$ and $R^2 R_B$, which can be taken as measures of how much
$\Psi$ and $b$ curve the spacetime, for three different values of $q$
close to the threshold $q_*$. In Fig.~\ref{fig:mixed-q-low}, with
$q=q_a\simeq 0.918$, the scalar field grows with $T$ while the
solution is approximately DSS, and its stress-energy content dominates
$R_B$, for $T > 3.5$, until both fields eventually disperse (as
$p<p_*$ in this evolution). In Fig.~\ref{fig:mixed-q-high}, with
$q=q_c\simeq 0.922$, the scalar field is decaying while the solution
is approximately DSS: the amplitude of $R^2 R_\Psi$ grows until $T
\simeq 4.5$ and then it decays while that of $R^2 R_B$ grows until $T
\simeq 6$, after which both fields disperse. In
Fig.~\ref{fig:mixed-q-middle}, with the intermediate value
$q=q_b\simeq 0.920$, both fields $\Psi$ and $B$ seem to stay at
approximately the same relative amplitude until they both disperse. It
is difficult to tell whether $\Psi$ grows or decays because the
interval where the solution is approximately DSS is short, and this
makes it harder to determine $q_*$ precisely. However, we are
confident that $q_a<q_*<q_c$, with $q_*\simeq q_b$ our best
approximation (for the Gaussian initial data). To improve the
bisection in $q$, one would need to run our time evolutions in
quadruple precision, so as to better fine-tune $p_*$ and thus observe
more echoing before the fields disperse or form a black hole. As that
is computationally much more time-consuming, we have not attempted it.

For comparison with
  Figs.~\ref{fig:mixed-q-low}-\ref{fig:mixed-q-high},
Fig.~\ref{fig:mixed-q-pure-cases-references} illustrates $R^2 R_\Psi$
for the pure scalar field critical solution $(q=0)$ and $R^2 R_B$ for
the pure gravitational wave critical solution $(q=1)$.

\begin{figure}[!ht]
\centering
\includegraphics[scale=0.7]{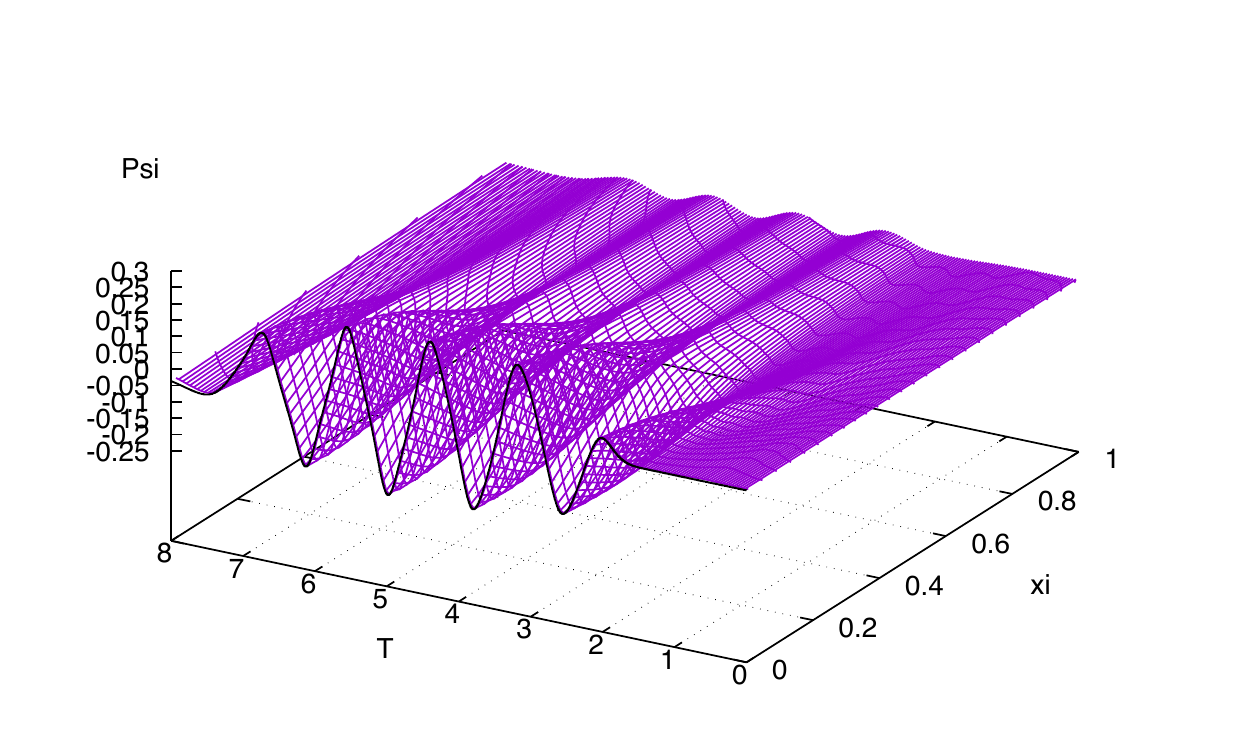}
\caption{The scalar field $\Psi(\xi, T)$ for optimal fine-tuning with
  $q=q_b$. A black line represents the extrapolation to the
  regular centre $R=0$. }
\label{fig:qc-psi}
\end{figure}

\begin{figure}[!ht]
\centering
\includegraphics[scale=0.85]{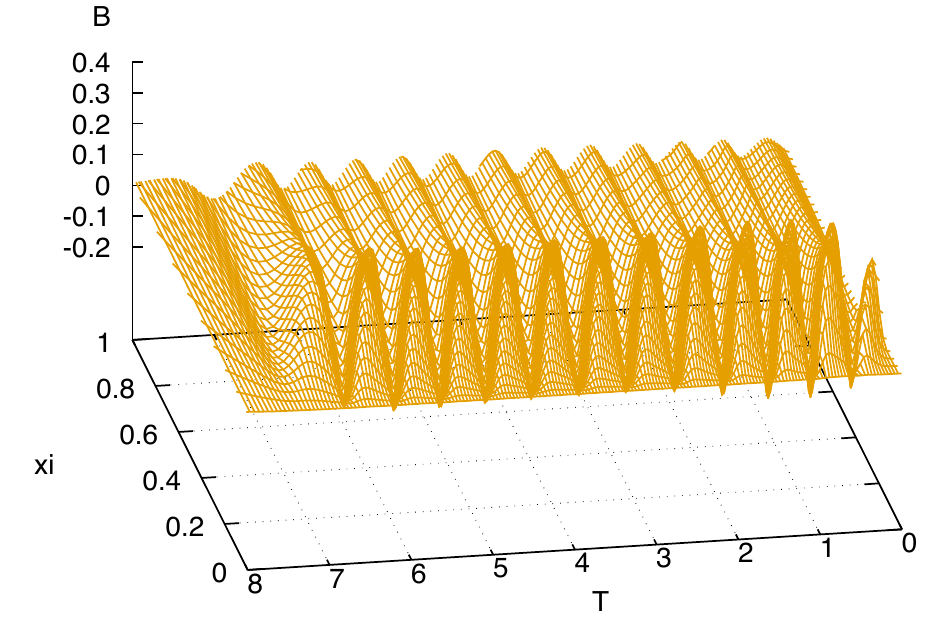}
\caption{The field $B$ for optimal fine-tuning with $q=q_b$. It is zero at
  the origin $R=0 \Leftrightarrow \xi=0$ due to
  Eq.~\eqref{eq:small-b}. }
\label{fig:qc-B}
\end{figure}

\begin{figure}[!ht]
\centering
\includegraphics[scale=0.8]{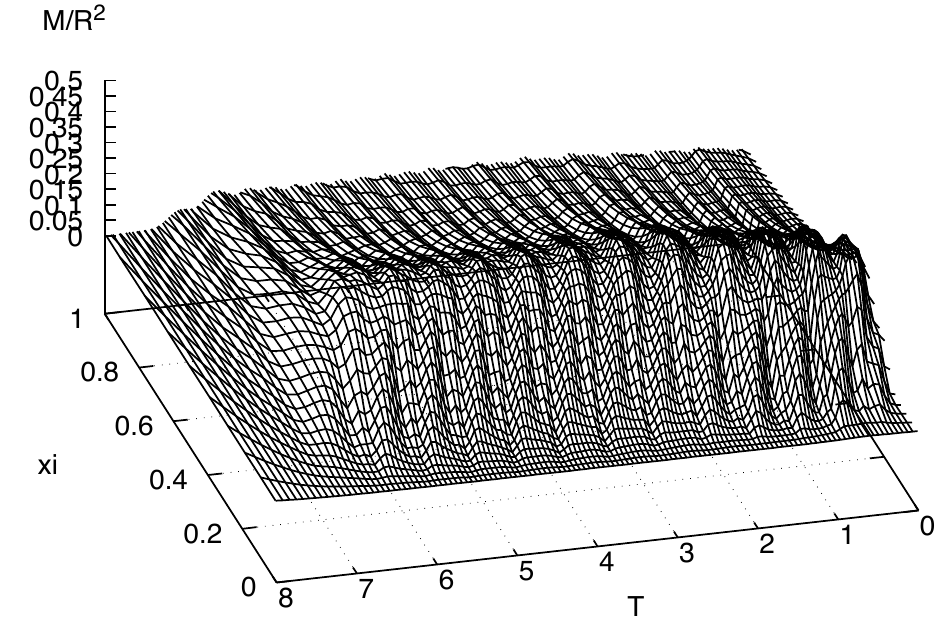}
\caption{The compactness $M/R^2$ for optimal fine-tuning with 
    $q=q_b$. }
\label{fig:qc-compactness}
\end{figure}

Figs.~\ref{fig:qc-psi}-\ref{fig:qc-compactness} show $\Psi$, $B$ and
$M/R^2$ for the best subcritical evolution with Gaussian initial data
and with $q=q_b$, which was our best estimate of $q_*$ up to two
decimal digits. We observe that $\Psi$ and $B$ are approximately
neither growing or decaying for $2.5 \leq T \leq 7$ before
dispersing.

From the data underlying these figures, we have estimated the echoing
periods of $\Psi$, $B$, ${\mathcal C}$, $R^2 R_\Psi$ and $R^2 R_B$ as
follows. We take discrete Fourier transforms of $(\max_x \Psi)(T)$
and $(\max_x B)(T)$ for a suitable interval of $T$, and adjust the
resulting period for what seemed the best fit by eye. Although this is
subjective, from the quality of the fit we estimate that we can
determine the periods within $\sim 0.01$. The results are given,
separately for $\Delta_\Psi/2$ and $\Delta_B$, and for different
$q\simeq q_*$, in Tables~\ref{table:DeltaGaussian} and
\ref{table:DeltaGaussianderiv}, respectively.

Although the separately fitted values of $\Delta_\Psi/2$ and
$\Delta_B$ are not equal, they are roughly within our estimate of the
accuracy $\sim 0.01$ to which we can determine these periods. Note
that the variation of the periods with $q$ over the ranges of $q$
considered in the table is somewhat larger than the difference of
$\Delta_\Psi/2$ and $\Delta_B$ at the same $q$. (As already discussed,
we are not able to determine $q_*$ very accurately.)

As further tests, we have also compared the fitted values of
$\Delta_\Psi/2$ and $\Delta_B$ to our plots of $R^2 R_\Psi$ and $R^2
R_B$, respectively, and find that they match well. Finally, we are
confident that $\Delta_B \lesssim \Delta_{\mathcal{C}}\lesssim \Delta_\Psi / 2$
(consistent with all being equal).

In short, our observations are consistent both with
$\Delta_B=\Delta_\Psi/2$ and $\Delta_B<\Delta_\Psi/2$. In other
words, we cannot decide if the critical solution is periodic (DSS) or only
quasiperiodic in $T$.

We note, however, that in the system for which this one is a
toy model, the Einstein-Maxwell equations in twistfree axisymmetry,
all fields in the critical solution are clearly only quasi-periodic
\cite{Baumgarte_Gundlach_Hilditch_2019}, already when viewed on their
own. By contrast, the quantities in
Figs.~\ref{fig:qc-psi}-\ref{fig:qc-compactness} seem, by eye, to be
periodic. One may take this to be an argument in favour of strict
DSS. 

\begin{table}[]
\centering
\caption{Estimated periods $\Delta_\Psi/2$ and $\Delta_B$ for Gaussian initial data.}
\begin{tabular}{l|l|l}
$q$            & $\Delta_\Psi / 2$ & $\Delta_B$ \\ \hline
0.9184570312 $=q_a$ & 0.61              & 0.59       \\
0.9200439452 $=q_b$ & 0.59              & 0.57       \\
0.9216308593 $=q_c$ & 0.5825              & 0.56       \\
0.9248046875 & 0.574              & 0.55      
\end{tabular}
\label{table:DeltaGaussian}
\end{table}

\begin{table}[]
\centering
\caption{Estimated periods $\Delta_\Psi/2$ and $\Delta_B$ for Gaussian derivative initial data.}
\begin{tabular}{l|l|l}
$q$      & $\Delta_\Psi / 2$ & $\Delta_B$ \\ \hline
0.859375  & 0.5875            & 0.55       \\
0.8671875 & 0.56              & 0.518      \\
0.87      & 0.55              & 0.512      \\
0.8725    & 0.55              & 0.511     
\end{tabular}
\label{table:DeltaGaussianderiv}
\end{table}

Recall that $\Delta_\Psi\simeq 1.6$ and $\Delta_B\simeq 0.47$ in the
pure scalar field and gravitational wave critical solutions,
respectively. So we can at least say that $\Delta_\Psi/2$ and
$\Delta_B$ have moved  from their pure values towards a common
intermediate value in the bi-critical solution.


\begin{figure}
\includegraphics[scale=0.62, angle=0]{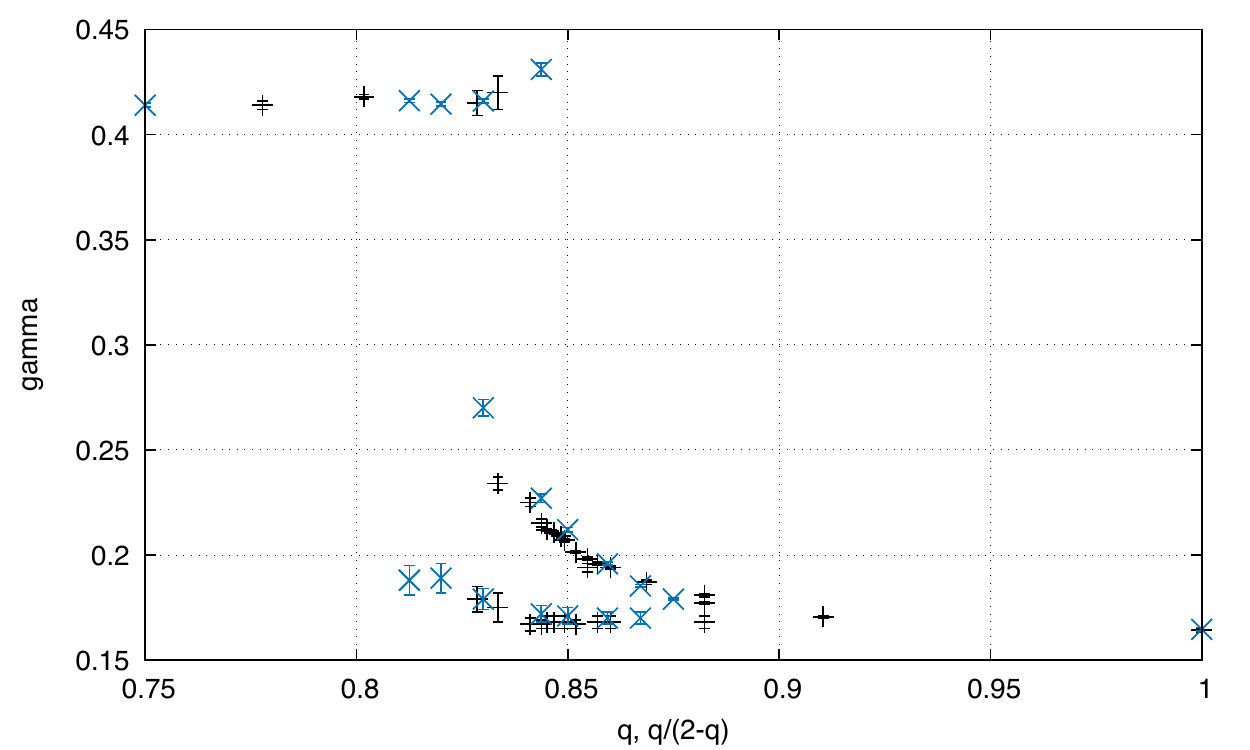} 
\caption{Plot of the critical exponent $\gamma$ estimated from the
  scaling law for radius of apparent horizon formation $R_\text{ah}$. The
  points in black correspond to initial data with a Gaussian profile,
  which are plotted against $q$. The points in blue correspond to
  initial data with the profile of a Gaussian derivative, which are
  plotted against $sq/(1-(1-s)q)$ for $s=0.5$.}
\label{fig:gamma-q}
\end{figure}

Fig.~\ref{fig:gamma-q} illustrates the estimated value of $\gamma$ for
different $q$, calculated from the scaling laws for the radius of
apparent horizon formation $R_\text{ah}$. To test universality, we
present the results for initial data with a Gaussian profile (in
black) and for initial data with the profile of a Gaussian derivative
(in blue). As $q_*$ depends on the family, the black points
are plotted against $q$, and the blue points are plotted against
\begin{equation}
    \tilde{q} := \frac{sq}{1-(1-s)q}
\end{equation}
with $0 \leq s \leq 1$ a free parameter. This transformation
has $q=0$ and $q=1$ as fixed points, with slope $1$ near $q=0$
  and slope $s$ near $q=1$. By adjusting $s$ we can ensure that the
neighbourhood around $q_*$ is located approximately at the same region
in the $\tilde q$-axis for both curves. We have set
$s=0.5$. 

From Fig.~\ref{fig:gamma-q} we see that for both our
  2-parameter families of initial data, $\gamma \simeq 0.41$ for
$q=0$, corresponding to the scalar field critical solution, and its
does not vary significantly with $q$ until $|q-q_*| \simeq 0.02$. In
this interval, the black hole mass scaling exponent depends on $\ln
(p-p_*)$: for poor fine-tuning, we find $\gamma \simeq 0.168$, close
to gravitational wave critical solution, and for better fine-tuning
its value is slightly higher and dependent on $q$, decreasing
monotonically from $\gamma \simeq 0.22$ to $\gamma \simeq
0.18$. This break in the scaling laws corresponds to the
  transition from a growing scalar field perturbation to the true
  bi-critical solution in near-critical time evolutions, as seen in
  Figs.~\ref{fig:qc-psi}-\ref{fig:qc-compactness}. As $q$ approaches
$1$, $\gamma$ settles to the value $\gamma \simeq 0.164$ of the
  gravitational wave critical solution. For this range of $q$, the
exponent is small, which is why the number of echoing periods seen is
limited when fine-tuning in $p$ up to double-precision.


\section{Conclusions}
\label{section:conclusions}


We have studied the threshold of black hole formation for a massless
scalar field minimally coupled to the gravitational wave metric ansatz
of \cite{Bizon_Chmaj_Schmidt_2005} in 4+1 dimensions
\cite{Bizon_Chmaj_Schmidt_2005}, (the latter restricted to the biaxial
case). We think of this as a toy model for matter gravitational
collapse beyond spherical symmetry, where gravitational waves are also
necessarily present.

We found that weak gravitational wave perturbations of the scalar
field critical solution decay, while weak scalar perturbations of the
gravitational wave critical solution also decay. This is different
from the case of critical collapse of two massless matter fields
\cite{Gundlach_Baumgarte_Hilditch_2019}, in which scalar perturbations
on the Yang-Mills field critical solution grow, but Yang-Mills
perturbations on the scalar field critical solution decay.

\begin{figure}
\includegraphics[scale=0.27, angle=0]{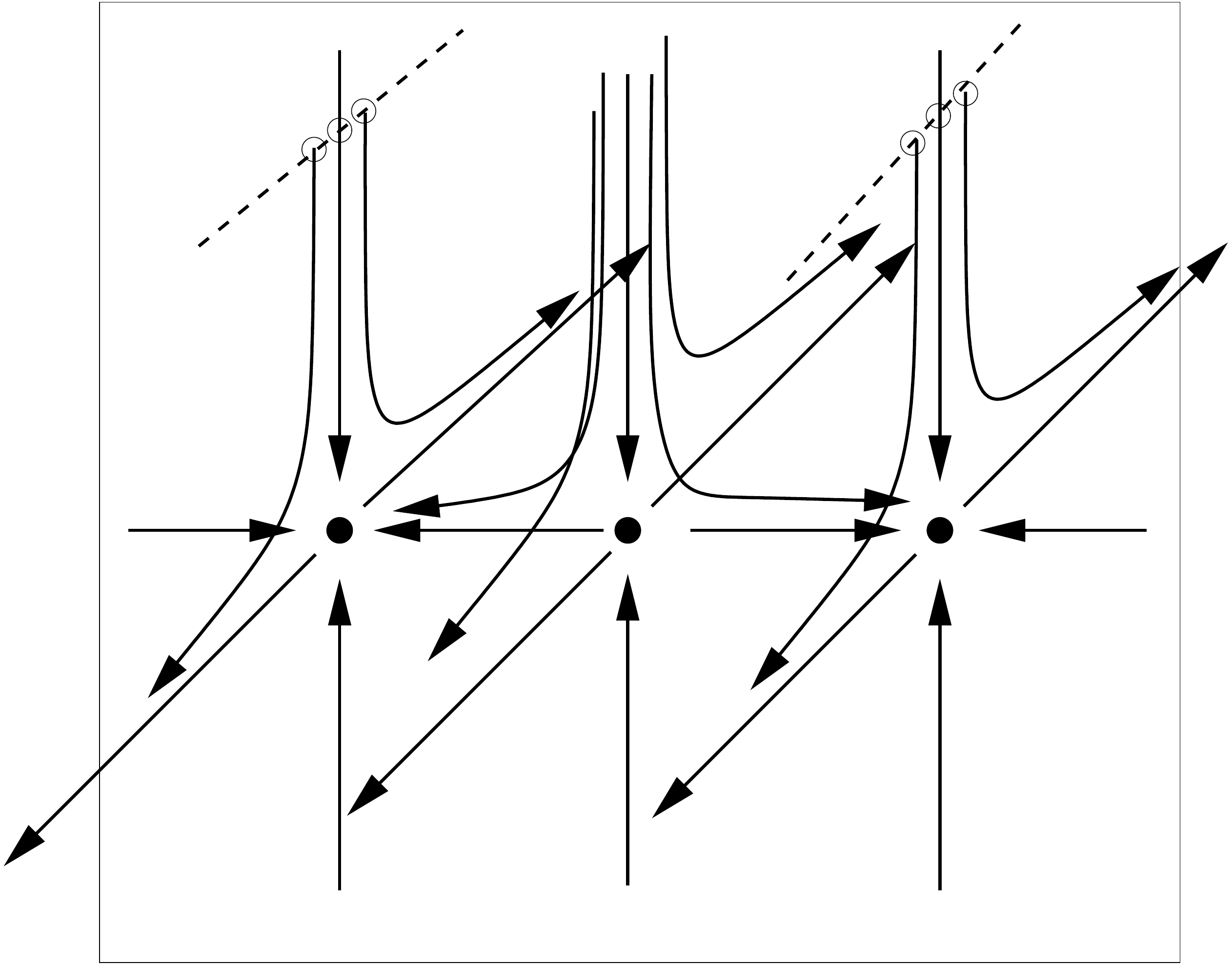} 
\caption{Schematic conjectured phase space picture, with the
  infinite-dimensional phase space represented in three
  dimensions. The framed plane represents the black hole threshold (in
  reality a hypersurface). All arrow lines represent trajectories
  (spacetimes). The filled dots represent fixed points (DSS
  spacetimes): the scalar field critical solution, on the left, the
  gravitational wave critical solution, on the right, and the
  codimension-two critical solution in between. Here the middle fixed
  point has two unstable modes, while the left and right ones have one
  each. An infinite number of phase space dimensions of the black hole
  threshold are suppressed, and with them an infinite number of stable
  modes of each fixed point within the black hole threshold. The two
  dashed lines represent three families of initial data with $q=0$
  (left) and $q=1$ (right). Hollow dots represent initial data with
  $p<p_*$, $p=p_*$ and $p>p_*$ for each family. Figure taken from
  \cite{Gundlach_Baumgarte_Hilditch_2019}.}
\label{fig:phasespace}
\end{figure}

These observations suggest the schematic phase space picture of
Fig.~\ref{fig:phasespace}. Here, any point in the phase space
represents an initial data set, up to an overall length scale,
parameterised in our case as $(\Psi(x),\chi(x))$, and a time evolution
curve corresponds to a spacetime, in our case in null slicing, again
up to an overall scale, with the time $T$ of the dynamical system
determining the missing scale as $e^{-T}$. In this picture, a DSS
solution should be a closed curve, but for simplicity we represent it
as a fixed point.

To find the bi-critical solution suggested by this picture, we then
explored the transition between the two pure critical solutions for
mixed initial data in our new toy model.
 
The evidence for the existence of the hypothetical codimension-two
attractor comes from the behaviour of our best near-critical [that is,
  $p\simeq p_*(q)$] evolutions for different values of $q$. In the
limit of perfect fine-tuning of $p$, as the mixing parameter $q$
decreases from $1$, we observe a transition from the gravitational
wave critical solution to the scalar field critical solution. By
continuity, we expect there to be a $q_*$ such that, in the limit of
perfect fine-tuning to $p=p_*(q_*)$, both fields play equal dynamical
roles. Increasing or decreasing $p$ an infinitesimal amount above or
below the curve $p=p_*(q)$ would push the critical solution to
eventual collapse or decay, respectively, while increasing or
decreasing $q$ exactly along this curve would push it into decaying
into the pure gravitational wave or pure scalar critical solutions, respectively.

The numerical limits of fine-tuning do not allow us to follow the
putative bi-critical solution for given $q$ down to arbitrarily large
$T$, but our observations are consistent with the assumption that in
the limit $(q,p)=(q_*, p_*(q_*))$, the system evolves toward an
intermediate attractor for which $\Psi$ and $B$ neither grow nor
decay.

Going beyond that, we want to know if the bi-critical solution is
strictly DSS, with a common period for all variables (in the sense that
$\Delta_\Psi=2\Delta_B$), or only quasiperiodic. Unfortunately,
because we observe the bi-critical solution over few periods,
Figs.~\ref{fig:qc-psi}-\ref{fig:qc-compactness} and
Fig.~\ref{fig:mixed-q-middle} seem to be compatible both with
$\Delta_\Psi/2=\Delta_B$ or with a slightly smaller value of $\Delta_B$.

With solutions of the toy model depending only on radius and time, one
might hope to construct a strictly DSS solution (as the
  hypothetical bi-critical solution) by ansatz, imposing periodic
boundary conditions in $T$ with a period $\Delta$ to be solved
for. Such an ansatz was solved numerically for the spherical scalar
field in 3+1 dimensions in \cite{Gundlach_Chop1}, and the numerical
approximate solution was leveraged into a proof of existence as a
real-analytic exact solution in \cite{ReiTru12}. However, a failure to
find an approximate numerical solution  of such an ansatz would
not prove the absence of an exact DSS solution, as the numerical
solution of a highly nonlinear boundary value problem may simply not
converge from an initial guess that is too rough. By contrast, it is
not clear how one could even make an ansatz of quasi-periodicity.


\appendix


\section{Scalar field equations in spherical symmetry in $n+2$
  dimensions}
\label{appendix:scalarfield}


In this Appendix, we explore the problem of a massless scalar field
minimally coupled to gravity in a spherically symmetric spacetime in
$n+2$ dimensions. We use coordinates $(u, x, \Omega_n)$, where $u$ and
$x$ are the same as defined in Section~\ref{sec:metric-and-equations},
and $\Omega_n$ are coordinates on the $n$-sphere:
\begin{equation}
    ds^2 = -2g R_{,x} dudx - Hdu^2 + R^2 d\Omega^2_n
\end{equation}
The Einstein equations 
\begin{equation}
R_{ab}=8\pi \nabla_a \Psi\nabla_b \Psi
\end{equation}
and the scalar field wave equation
\begin{equation}
    \nabla^a \nabla_a \Psi = 0
\end{equation}
can be put in the following hierarchy in these coordinates:
\begin{align}
  \mathcal{D}(\ln g)&= \frac{8\pi R}{n} (\mathcal{D} \Psi )^2, \\
  \mathcal{D}(R^{n-1} \Xi R) &= -\frac{n-1}{2} g R^{n-2},
  \\
  \mathcal{D}(R^{n/2} \Xi \Psi)&= -\frac{n}{2}  R^{n/2-1} \Xi R
  \mathcal{D} \Psi.
\end{align}
Using boundary conditions at $R=0$, we write the above
equations in integral form to make the link to the numerical
integrations more explicit:
\begin{align}
\label{eq:appendix-g-tmp}
 g &= \exp\left[\frac{4\pi }{n}\int_0^R  (\mathcal{D} \Psi )^2
 \, d(\tilde{R}^2)\right],\\
 \label{eq:appendix-xi-r}
    \Xi R &= -\frac{1}{2} \frac{1}{R^{n-1}} \int_0^R g \, d(\tilde{R}^{n-1}),\\
\label{eq:appendix-xi-psi}
    \Xi \Psi &= -\frac{1}{R^{n/2}} \int_0^R 
  \mathcal{D} \Psi \Xi R \, d(\tilde{R}^{n/2}).
\end{align}


\begin{figure}[!ht]
\centering \includegraphics[scale=0.7]{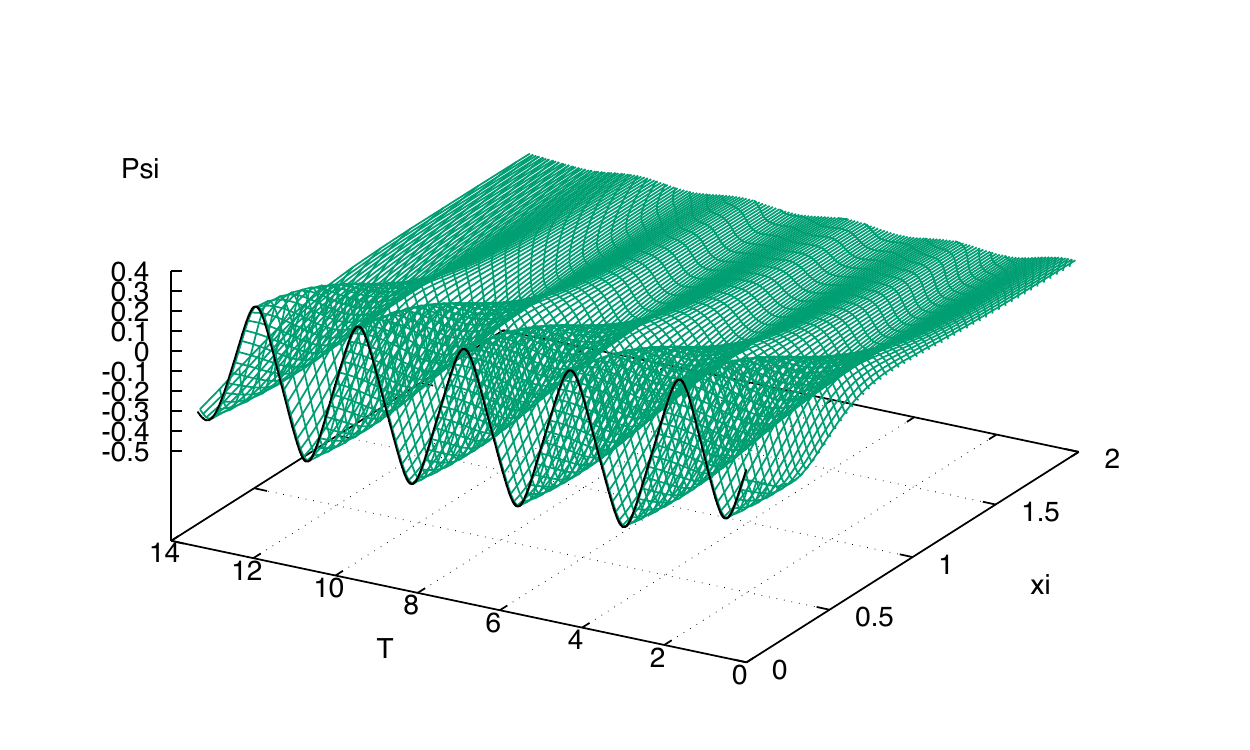}
\caption{The scalar field $\Psi(x, T)$ in the best near-critical
  evolution in 8+1 dimensions. A black line represents the extrapolation
  to the regular centre $R=0$. }
\label{fig:9dim-psi}
\end{figure}

\begin{figure}[!ht]
\centering \includegraphics[scale=0.7]{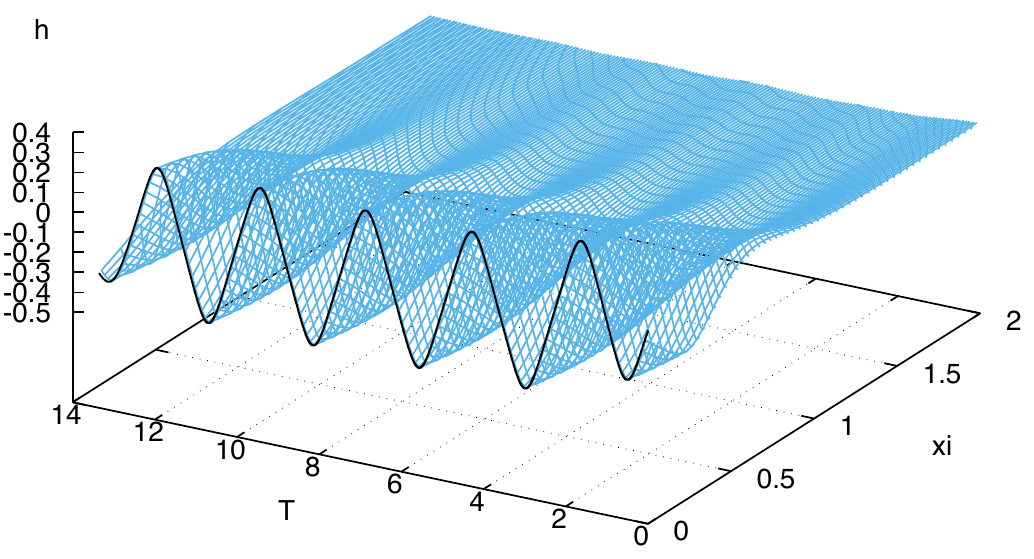}
\caption{The field $h(x, T)$ in the best near-critical evolution in 8+1
  dimensions. A black line represents the extrapolation to the regular
  centre $R=0$. }
\label{fig:9dim-h}
\end{figure}

\begin{figure}[!ht]
\centering
\includegraphics[scale=0.8]{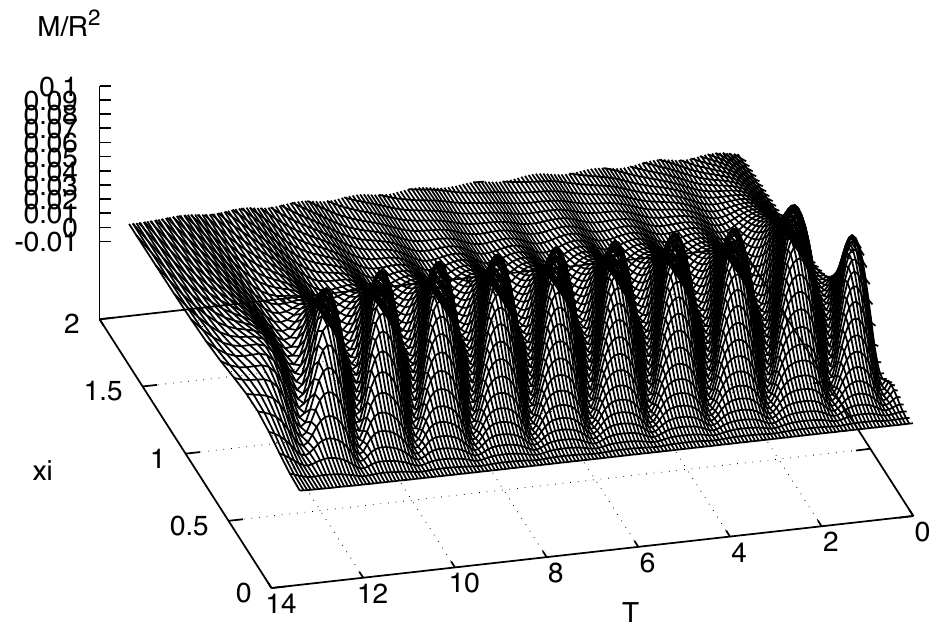}
\caption{The compactness $M/R^2$ in the best near-critical evolution
  in 8+1 dimensions. A black line represents the extrapolation to the
  regular centre $R=0$.}
\label{fig:9dim-compactness}
\end{figure}


The division by $R^{n-1}$ to calculate $\Xi R$ in
Eq.~\eqref{eq:appendix-xi-r} generates numerical instabilities near
the origin $R=0$ when the dimension increases. While it produces no
significant effect in $n+2 \leq 5$ dimensions, in 8+1 dimensions it
leads to unphysical behavior in $\Xi R$. A simple solution to this is
to integrate the equation by parts as suggested in
\cite{Bland_Preston_Becker_Kunstatter_Husain_2005} and to use
Eq.~\eqref{eq:appendix-g-tmp}:
\begin{equation}
 \label{eq:new_gbar}
\Xi R = \frac{g}{2} + \frac{4\pi}{n(n+1)}\frac{1}{R^{n-1}} 
\int_0^R g (\mathcal{D} \Psi )^2 \, d(\tilde{R}^n)
\end{equation}
The second term in Eq.~\eqref{eq:new_gbar} can be more accurately
computed as it is $O(R^2)$ near the origin.

In a similar manner, the wave equation,
Eq.~\eqref{eq:appendix-xi-psi}, displays instabilities in 8+1 dimensions
near the origin $R=0$ which arise from integrating over and dividing
by $R^{n/2}$ the term on the right hand side, which is $O(1)$ for
small $R$. To avoid this, we define a field $h$ as
\begin{equation}
    \label{eq:appendix-h-from-psi}
    h := \frac{d(R^{n/2}\Psi)}{d(R^{n/2})}= \Psi 
+ \frac{2}{n}(\mathcal{D} \Psi)R, 
\end{equation}
from which we can recover
\begin{equation}
    \label{eq:appendix-psi-from-h}
    \Psi = \frac{1}{R^{n/2}}\int_0^R
h \, d(\tilde{R}^{n/2}).
\end{equation}
 (An evolution equation for $h$ follows below). The problematic
integral in Eq.~\eqref{eq:appendix-xi-psi}, which is no longer needed,
appears to have simply been replaced by another problematic integral,
Eq.~(\ref{eq:appendix-psi-from-h}).  However, this can again be
  integrated by parts to make it more explicitly regular, whereas
  integration by parts would not be useful for
  Eq.~\eqref{eq:appendix-xi-psi}.

The  final form of our field equations can be collected in
the following hierarchy:
\begin{align}
\label{eq:appendix-psi-from-h-by-parts}
\Psi &= h -\frac{1}{\frac{n}{2}+1}\frac{1}{R^{n/2}} \int_0^R
\mathcal{D} h \, d(\tilde{R}^{n/2+1}), \\ 
\label{eq:appendix-g}
g &= \exp\left[\int_0^R 2\pi n
  \frac{(h-\Psi)^2}{\tilde{R}} \, d(\tilde{R})\right],\\ 
\Xi R &=
\frac{g}{2} + \frac{4\pi}{n(n+1)}\frac{1}{R^{n-1}} \int_0^R g
(\mathcal{D} \Psi )^2 \, d(\tilde{R}^n),\\
\label{eq:appendix-xi-h}
\Xi h &= \frac{1}{2R}(h-\Psi)\left[(n-1)g+\frac{n}{4}\Xi R \right].
\end{align}
The second term on the right hand side of
Eq.~\eqref{eq:appendix-psi-from-h-by-parts} below is $O(R)$ at the
origin, and thus more stable to compute than $\Xi \Psi$. [It is not
  useful to integrate the expression for $\Xi \Psi$ in
  Eq.~\eqref{eq:appendix-xi-psi} by parts directly, as the integrand
  would involve second-order derivatives of $\Psi$.]  The new
evolution equation (\ref{eq:appendix-xi-h}) does not require an
integral and does not come with high powers of $R$.
Eqs.~(\ref{eq:appendix-g}) and \eqref{eq:appendix-xi-h} are well
defined at the origin as $h-\Psi = O(R)$ by
Eq.~\eqref{eq:appendix-h-from-psi} and by regularity of $\Psi$. In
$n+2=4$ dimensions in particular, it is $O(R^2)$ and reduces to $\Xi
h=0$ in Minkowski spacetime, where $g=-\Xi R/2=1$.

This field transformation has been commonly used in 4 dimensions, for
example in
\cite{Goldwirth_Piran_1987,Gundlach_Price_Pullin_1994,Garfinkle_1995}.
In \cite{Garfinkle_Cutler_Comer_Duncan_1999},  Garfinkle~{\it
et~al.} introduced a generalization of $h$ from 3+1 to higher
spacetime dimensions, completely different from
Eq.~\eqref{eq:appendix-psi-from-h}, that maintains  the property
of $h$ being constant along ingoing light rays in Minkowski
spacetime, $\Xi h=0$. This  is possible only for even $n$, as
solutions of the wave equation in flat spacetime satisfy Huygens’
principle only in even spacetime dimensions. We have tried to explain
in this Appendix why the definition of $h$ of  Bland~{\it
et~al.} \cite{Bland_Preston_Becker_Kunstatter_Husain_2005} is numerically
advantageous even though for $n\ne 2$ it does not have the 
very property that  seems to have motivated its introduction
in $n=2$.

As an indication that our implementation of this formulation works,
Figs.~\ref{fig:9dim-psi}-\ref{fig:9dim-compactness} show the critical
solution in 8+1-dimensional spherical scalar field collapse, 
found by fine-tuning the amplitude of a family of initial data to
the collapse threshold.

In the main paper, we are concerned with the dynamics of the field
$b$, whose governing equation is mathematically similar to that of the
scalar wave equation in 8+1 dimensions. As in odd spacetime dimensions
we cannot use the methods of
\cite{Garfinkle_Cutler_Comer_Duncan_1999}, we have adopted the
formulation described here for arbitrary integer $n$, with our $\chi$
and $b$ in 4+1 dimensions the equivalents of $h$ and $\Psi$ in 8+1
dimensions.

\vfill


\bibliography{references}


\end{document}